\pdfoutput=1
\newcommand*{\ATLASLATEXPATH}{}
\documentclass[PAPER, cernpreprint, atlasdraft=true, texlive=2016, UKenglish]{\ATLASLATEXPATH atlasdoc}
 
\usepackage{\ATLASLATEXPATH atlaspackage}
\usepackage{\ATLASLATEXPATH atlasbiblatex}
 
\usepackage[hepprocess, hepparticle]{\ATLASLATEXPATH atlasphysics}
\usepackage{multirow}
 
\graphicspath{{logos/}{figures/}}
 
\usepackage{ANA-HIGG-2018-14-PAPER-defs}

\AtlasTitle{Test of CP invariance in vector-boson fusion production of the Higgs boson
in the $H\rightarrow\tau\tau$ channel
in proton--proton collisions at $\sqrt{s} = 13\,\text{Te\kern -0.1em V}$ with the ATLAS detector}
 
\AtlasJournal{Phys.\ Lett.\ B.}
\AtlasJournalRef{Phys.\ Lett.\ B.\ 805\ (2020)\ 135426}
\AtlasDOI{10.1016/j.physletb.2020.135426}
\AtlasAbstract{
A test of CP invariance in Higgs boson production via vector-boson fusion
is performed in the $H\rightarrow\tau\tau$ decay channel. This test uses the Optimal Observable
method and is carried out using
$36.1\,\mathrm{fb}^{-1}$ of $\sqrt{s}=13\,\text{Te\kern -0.1em V}$ proton--proton collision data collected by the ATLAS
experiment at the LHC. Contributions from CP-violating interactions between the Higgs boson and electroweak
gauge bosons are described by an effective field theory,
in which the parameter $\tilde{d}$~governs the strength of CP violation.
No sign of CP violation is observed in the distributions of the Optimal Observable,
and $\tilde{d}$~is constrained to the interval
[$-0.090$,\,$0.035$] at the 68\% confidence level (CL), compared to an expected
interval of $\tilde{d} \in [-0.035,0.033]$ based upon the Standard Model prediction.
No constraints can be set on $\tilde{d}$ at 95\% CL,
while an expected 95\% CL interval of $\tilde{d} \in [-0.21,0.15]$ for the Standard Model hypothesis
was expected.
}
 
\author{The ATLAS Collaboration}
 
\PreprintIdNumber{CERN-EP-2020-009}

\AtlasRefCode{HIGG-2018-14}

\hypersetup{pdftitle={ATLAS document},pdfauthor={The ATLAS Collaboration}}
 
\begin{document}
 
\maketitle
 
\tableofcontents

\section{Introduction}
One of the central puzzles in physics today is the observed baryon asymmetry of the universe.
The violation of invariance of fundamental interactions under the transformation of charge
conjugation (C)  and its combination with parity (CP) is one of the three necessary Sakharov
conditions~\cite{Sakharov:1967dj} to explain the dynamical generation of the baryon asymmetry.
In the Standard Model (SM) of particle physics, CP violation (CPV) is introduced via
the complex phase in the quark mixing (CKM) matrix~\cite{Cabibbo:1963yz,Kobayashi:1973fv}.\footnote{Effects of possible CPV in the neutrino sector and in the strong interaction
are not considered in this statement.} It is able to describe
all observations of CPV in the $K$-, $B$-, and $D$-meson systems~\cite{Christenson:1964fg,AlaviHarati:1999xp,Lai:2001ki,
Aubert:2001nu,Abe:2001xe,Aubert:2004qm,Chao:2004mn,Aaij:2013iua,Aaij:2012kz,Aaij:2012kz,Aaij:2019kcg,Charles:2015gya,Alpigiani:2017lpj}.
However,  the measured size of the complex phase and the derived magnitude of CPV in the early universe are insufficient to
explain the observed value of the baryon asymmetry within
the SM~\cite{Huet:1994jb,Gavela:1993ts,Cohen:1993nk,Riotto:1999yt,Hou:2008xd} and, most probably,
new sources of CPV beyond the SM need to be introduced.
 
The investigation of Higgs boson production and decay at the LHC
offers a novel opportunity to search for
new sources of CPV in the interaction of the Higgs boson with other SM particles.
No observable effect of CPV is expected in the production or decay of the SM Higgs boson. Hence any observation of
CP violation involving the observed Higgs boson ~\cite{HIGG-2012-27,CMS-HIG-12-028}  would be an unequivocal sign of physics
beyond the SM.
 
The measured Higgs boson production cross sections, branching ratios, and derived constraints on coupling-strength modifiers,
assuming the tensor structure of the SM, agree with the SM predictions
within the current precision~\cite{HIGG-2015-07,Aad:2019mbh,CMS-HIG-17-031}.
Investigations of spin and CP quantum numbers strongly indicate that the observed particle is of scalar nature and that
the dominant coupling structure is CP-even and consistent with the SM expectation~\cite{HIGG-2013-01,HIGG-2013-17,CMS-HIG-12-041}.
Various measurements have been used in the framework of effective field theories to derive limits on
Wilson coefficients which multiply CP-even and CP-odd operators and modify the structure and strength of
the coupling of the Higgs boson to gluons and electroweak gauge bosons.  These include
measurements of differential cross sections as functions of CP-even observables in the decay  $H \rightarrow \gamma \gamma$~\cite{HIGG-2016-21},
measurements of event rates in specific event categories and phase-space regions in the decay $H \rightarrow ZZ^{\ast}$ ~\cite{HIGG-2016-22},
and measurements of the $VH$ invariant mass in Higgs boson production in association with a weak gauge boson $V$ ($V=W ^\pm, Z$) ~\cite{CMS-HIG-14-035}.
These analyses use CP-even observables and event rate information and hence are not directly sensitive to possible
interference between the CP-even SM operators and new CP-odd operators. The shapes of distributions of CP-odd and
CP-even observables (without exploiting CP-even rate information) have been used to set limits on CP-odd and CP-even couplings of the Higgs boson to gauge bosons. This is done by investigating the decay
$H \rightarrow V V^{\ast} (V=W^\pm,Z)$, using only information from the decay~\cite{HIGG-2013-17,CMS-HIG-14-018} and
combining it with information from vector-boson fusion (VBF) or associated $VH$ production
~\cite{CMS-HIG-17-011,CMS-HIG-18-002}. Another analysis using the decay $H \rightarrow \tau \tau$ exploits information from
VBF and $VH$ production~\cite{CMS-HIG-17-034}. The shape of the distribution of a single CP-odd observable
constructed from kinematic information in VBF production in $H \rightarrow \tau \tau$ candidate events has been
previously used to set a limit on the parameter $\dtilde$~\cite{HIGG-2015-06}, which governs
the strength of CPV in an effective field theory ansatz as described in Section~\ref{sec:theoryandmethodology}.
This analysis constrained $\dtilde$ to the interval $[-0.11,0.05]$ at the 68\% confidence level (CL) using ATLAS data collected
at $\sqrt{s}= 8$~\TeV\ in 2012, while a 68\% CL interval of $\dtilde \in [-0.16,0.16]$ was expected.
No hints of CPV have been observed in these studies.
 
In this Letter, a direct test of CP invariance
in Higgs boson production via VBF is presented in the $H \to  \tau \tau$ channel,
based on proton--proton collision data corresponding to an integrated luminosity of 36.1\,\ifb\ collected with the
ATLAS detector at $\sqrt{s}$ = 13~\TeV\ in the years 2015 and 2016. A CP-odd \OOtext~\cite{Atwood:1991ka, Davier:1992nw, Diehl:1993br} is employed. The \OOtext
combines the information from the multidimensional phase space in a single quantity calculated from
leading-order matrix elements for VBF production, independent of the decay mode of the Higgs boson.
VBF production provides a promising physics process to test CP invariance in the $HVV$ vertex~\cite{Brehmer:2017lrt}. The decay mode $H \to  \tau \tau$ allows
the selection of signal events with a good signal-to-background ratio and the reconstruction of the four-momentum of the
Higgs boson candidate with adequate precision.
 
In the present work a direct test of CP invariance is obtained through a measurement of the mean value of the CP-odd \OOtext, neglecting
possible effects from rescattering by new light particles in loops~\cite{Brehmer:2017lrt}. A measurement of the parameter $\dtilde$
is also performed.
Limits on $\dtilde$ are derived by analysing the shapes of distributions of the \OOtext measured in \Htt\ candidate events with two jets
in the final state consistent with VBF production. The event selection,
estimation of background contributions, and
systematic uncertainties closely follow the analysis employed for the observation of the \Htt\ decay~\cite{HIGG-2017-07}.  In order
to increase the signal-to-background ratio, the final event selection utilizes multivariate discriminants.
 
 
\section{Theoretical framework and methodology}
\label{sec:theoryandmethodology}
The effective Lagrangian ${\cal L}_{\mathrm{eff}}$ considered is the SM Lagrangian augmented with CP-odd operators of mass dimension six, involving the Higgs field
and electroweak gauge fields. No CP-even dimension-six operators built from these fields are taken into account.  All interactions between the Higgs boson and
other SM particles (fermions and gluons) are assumed to be as predicted in the SM, i.e. the coupling structure in gluon--gluon
fusion production and in the decay into a pair of $\tau$-leptons is considered to be the same as in the SM. The theoretical ansatz considered
and the methodology
is the same as in Ref.~\cite{HIGG-2015-06}, which contains further details.
After electroweak symmetry breaking, the Lagrangian can be written in the mass basis of the Higgs boson $H$, photon $A$ and weak gauge bosons
$W^\pm$ and $Z$ as in Ref.~\cite{Achard:2004kn}:
\begin{linenomath}
\begin{equation*}
{\cal L}_{\mathrm{eff}} = {\cal L}_{\mathrm{SM}} + \tilde{g}_{HAA} H \tilde{A}_{\mu\nu}{A}^{\mu\nu} +
\tilde{g}_{HAZ} H \tilde{A}_{\mu\nu}{Z}^{\mu\nu} +
\tilde{g}_{HZZ} H \tilde{Z}_{\mu\nu}{Z}^{\mu\nu} +
\tilde{g}_{HWW} H \tilde{W}^+_{\mu\nu}{W}^{-\mu\nu} \, ,
\end{equation*}
\end{linenomath}
where ${V}^{\mu\nu}$ and $\tilde{V}^{\mu\nu} = \epsilon^{\mu\nu\rho\sigma} {V}_{\rho\sigma}$
(with $V=W^\pm,Z, A$) denote the field strength and dual field strength tensors, respectively.
Only two of the four couplings $\tilde{g}_{HVV^\prime}$ are independent due to constraints imposed
by U(1)$_Y$ and SU(2)$_{I_W,{\mathrm L}}$ invariance. They can be expressed in terms of two dimensionless
couplings $\tilde{d}$ and $\tilde{d}_B$ as in Refs.~\cite{Buchmuller:1985jz,Hankele:2006ma}:
\begin{linenomath}
\begin{eqnarray*}
\tilde{g}_{HAA}  =   \frac{g}{2 m_W} (\tilde{d} \sin^2 \theta_W + \tilde{d}_B \cos^2 \theta_W) & &
\tilde{g}_{HAZ}  =   \frac{g}{2 m_W} \sin 2\theta_W (\tilde{d} -\tilde{d}_B)\\
\tilde{g}_{HZZ}  =  \frac{g}{2 m_W} (\tilde{d} \cos^2 \theta_W + \tilde{d}_B \sin^2 \theta_W) & &
\tilde{g}_{HWW}  =   \frac{g}{m_W} \tilde{d} \, ,
\end{eqnarray*}\end{linenomath}
where $g$ is the SU(2) coupling constant and $\theta_W$ is the weak mixing angle.
Adopting the arbitrary choice $\tilde{d} = \tilde{d}_B$ yields the following relations:\footnote{The parameter $\tilde{d}$ is related to the parameter $\hat{\kappa}_W= (\tilde{\kappa}_W/ \kappa_{\mathrm{SM}}) \tan \alpha$ used in the investigation
of CP properties in the decay $\HWW$
via $\tilde{d} = - \hat{\kappa}_W = - (\tilde{\kappa}_W/ \kappa_{\mathrm{SM}}) \tan \alpha$. The choice $\tilde{d} = \tilde{d}_B$ yields $\hat{\kappa}_W = \hat{\kappa}_Z$
as assumed in the combination of the $\HWW$ and $H\rightarrow ZZ^{\ast}$ decay analyses~\cite{HIGG-2013-17}.}
\begin{linenomath}\begin{equation*}
\tilde{g}_{HAA} = \tilde{g}_{HZZ} = \frac{1}{2} \tilde{g}_{HWW} = \frac{g}{2 m_W} \tilde{d}~~~~
{\mathrm{and}} ~~~~~~ \tilde{g}_{HAZ} = 0 \, .
\end{equation*}\end{linenomath}
In an effective field theory (EFT), the coupling parameters are real valued. However, rescattering effects from new particles
in loops, with masses lower than the scale of new physics assumed in the EFT, may introduce an imaginary part \cite{Brehmer:2017lrt}.
Such effects are not considered in the analysis presented here, as $\tilde{d}$ is assumed to be real valued.
 
The strength of CP violation in VBF Higgs boson production is then described by a single parameter $\tilde{d}$.
The corresponding matrix element $\mathcal{M}$ for VBF production is the sum of a CP-even contribution $\mathcal{M}_{\mathrm{SM}}$
from the SM and a CP-odd contribution $\mathcal{M}_{\textrm{CP-odd}}$ from the dimension-six operators considered:
\begin{linenomath}\begin{equation*}
\mathcal{M} =\mathcal{M}_{\mathrm{SM}}+\dtilde\cdot\mathcal{M}_{\textrm{CP-odd}},
\end{equation*}\end{linenomath}
where the dependence on $\dtilde$ has explicitly been factored out.
The squared matrix element has three contributions:
\begin{linenomath}\begin{equation*}
|\mathcal{M}|^{2}=|\mathcal{M}_{\mathrm{SM}}|^{2}+
\dtilde\cdot2 \operatorname{Re}(\mathcal{M}_{\mathrm{SM}}^{*}\mathcal{M}_{\textrm{CP-odd}}) +  \dtilde^{2}\cdot|\mathcal{M}_{\textrm{CP-odd}}|^{2} \, .
\end{equation*}\end{linenomath}
The first term $|\mathcal{M}_{\mathrm{SM}}|^{2}$ and third term  $\dtilde^{2}\cdot|\mathcal{M}_{\textrm{CP-odd}}|^{2}$ are both CP-even and hence are not a source
of CPV. The second term  $\dtilde\cdot2 \operatorname{Re}(\mathcal{M}_{\mathrm{SM}}^{*}\mathcal{M}_{\textrm{CP-odd}})$ stems
from the interference of the two contributions to the matrix element and is CP-odd, representing a possible new source of CPV in the Higgs sector.
The interference term integrated over a CP-symmetric part of phase space vanishes and therefore does not contribute to the total cross section and observed event yield after
CP-symmetric selection criteria are applied. The third term increases the total cross section by an amount quadratic in $\dtilde$, but this is not exploited in the analysis presented here as the observed rate can also be influenced by additional CP-conserving new physics.
 
The final state consisting of the reconstructed decay of the Higgs boson and the two tagging jets corresponding to the VBF topology
can be characterized by seven phase-space variables,  by fixing the mass of the Higgs boson, neglecting jet masses,
and exploiting momentum conservation in the plane transverse to the beam line. The concept of the \OOtext ($\mathcal{O}_\mathrm{opt}$)
combines the information from the seven-dimensional phase space into a single observable, which is shown to have the highest sensitivity
to small values of the parameter of interest and neglects contributions proportional to $\dtilde^2$ in the matrix element.
 
The \OOtext~for the determination of $\tilde{d}$ is given by the ratio of the interference term in the matrix element to the
SM contribution:
\begin{linenomath}\begin{equation*}
\mathcal{O}_\mathrm{opt} =\frac{2 \operatorname{Re}(\mathcal{M}_{\mathrm{SM}}^{*}\mathcal{M}_{\textrm{CP-odd}})}{|\mathcal{M}_{\mathrm{SM}}|^{2}} \, .
\end{equation*}\end{linenomath}
 
In order to make an almost model-independent test of CP invariance, the mean value of the \OOtext~
can be measured. If no CPV is present in the $HVV$ vertex, then the expectation value of the \OOtext~vanishes:
$\langle \mathcal{O}_\mathrm{opt} \rangle = 0$, as the \OOtext~is a CP-odd (and $\hat{\mathrm{T}}$-odd\footnote{$\hat{\mathrm{T}}$ denotes the naive time reversal according to Ref.~\cite{Brehmer:2017lrt}, which inverts the directions of momenta and spins.}) variable.  Since the initial state
of VBF production of the Higgs boson is not CP-symmetric, this argument assumes that effects from rescattering
are negligible~\cite{Brehmer:2017lrt}. Thus an observation of
a non-vanishing mean value or an asymmetry in the \OOtext~distribution would indicate physics beyond the SM,
either stemming from CPV, or originating from rescattering effects (i.e.\ new particles being on the mass shell in loop
corrections to the $HVV$ vertex).
Example distributions of the \OOtext~for signal events after the full event selection, as described in
Section\,\ref{sec:analysis}, are shown for various values of $\dtilde$  in Figure~\ref{figure:ootobs}. In the SM the
distribution is symmetric and has a mean value of zero, whereas a non-vanishing value of $\dtilde$ causes an asymmetry
and a non-vanishing mean value of the \OOtext.
\begin{figure}[h!b!t!]
\centering
\includegraphics[width=0.66\textwidth]{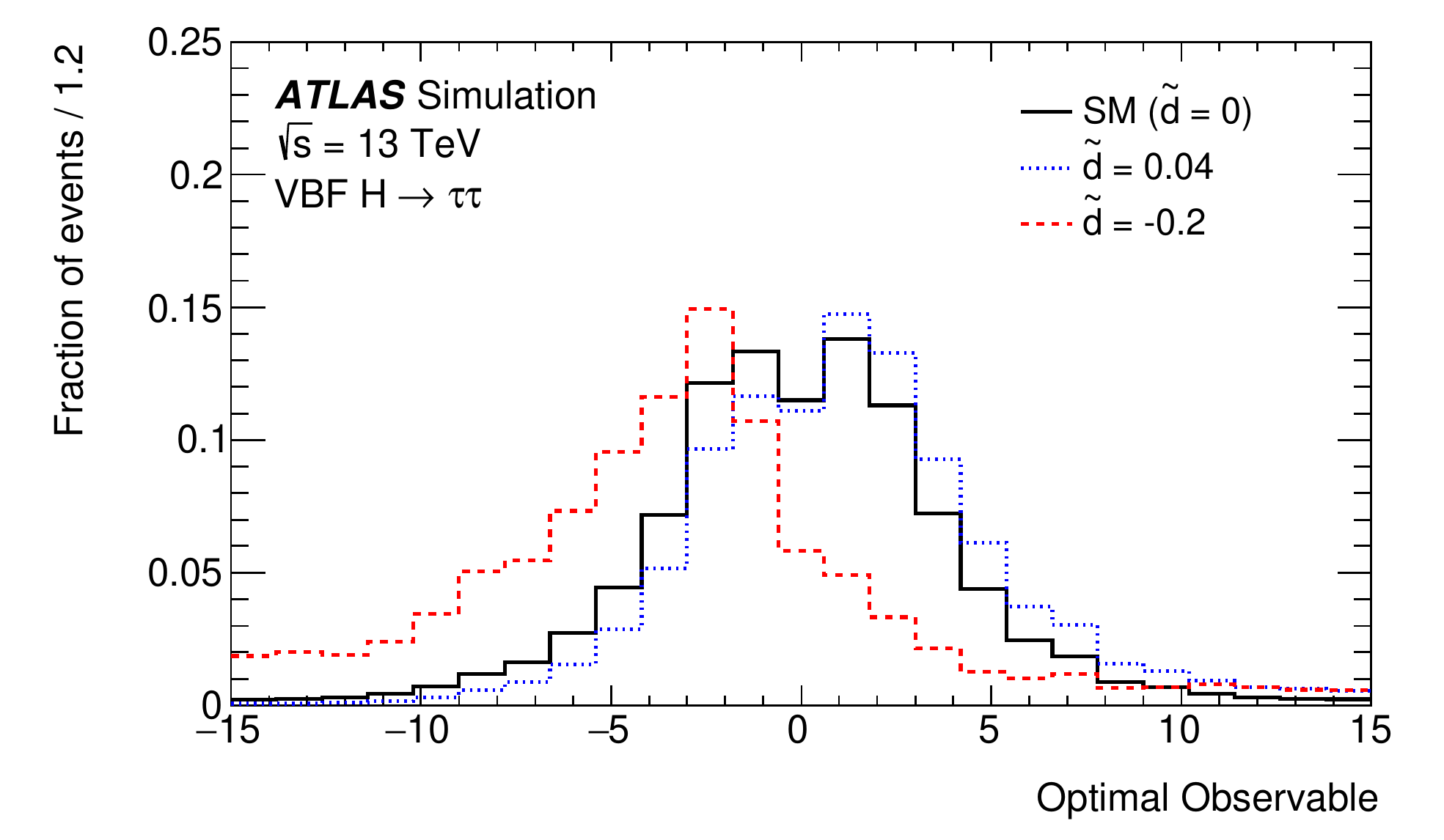}
\caption{Distribution of the \OOtext~for signal events for three example values of $\dtilde$
after event reconstruction and application of the full event selection used to define the signal region (see Section~\ref{sec:analysis}).
Non-vanishing values of $\dtilde$ cause an asymmetry and a non-vanishing mean value.}
\label{figure:ootobs}
\end{figure}
 
The values of the leading-order matrix elements (ME) needed for the calculation of the \OOtext are
extracted from HAWK~\cite{Ciccolini:2007jr,Ciccolini:2007ec,Denner:2014cla}. The evaluation requires the four-momenta of the Higgs boson and the two
tagging jets~($jj$).  The momentum fraction $x_{1}$ ($x_{2}$) of the initial-state parton from the proton moving in the positive (negative)
$z$-direction (along the beam) can be derived
by exploiting energy--momentum conservation from the Higgs boson and tagging jet four-momenta as
\begin{linenomath}\begin{equation*}
x_{1,2}^{\mathrm{reco}} = \frac{m_{Hjj}}{\sqrt{s}}\mathrm{e}^{\pm y_{Hjj}} ,
\end{equation*}\end{linenomath}
where $m_{Hjj}$ ($y_{Hjj}$) is the invariant mass (rapidity) obtained from the vectorially summed four-momenta of the
tagging jets and the Higgs boson.
Since the flavour of the initial- and final-state partons cannot be determined experimentally,
the sum over all possible flavour configurations $ij \rightarrow kl H$
weighted by the CT10 leading-order parton distribution functions (PDFs)~\cite{Lai:2010vv}
is calculated separately for the matrix elements in the numerator and denominator:
\begin{linenomath}\begin{eqnarray*}
2 \operatorname{Re}(\mathcal{M}_{\mathrm{SM}}^{*}\mathcal{M}_{\textrm{CP-odd}}) &=& \sum_{i,j,k,l}
f_i(x_1) f_j(x_2) 2 \operatorname{Re}((\mathcal{M}_{\mathrm{SM}}^{ij \rightarrow kl H})^*\mathcal{M}_{\textrm{CP-odd}}^{ij \rightarrow kl H})\\
|\mathcal{M}_{\mathrm{SM}}|^2 &=&  \sum_{i,j,k,l} f_i(x_1) f_j(x_2) |\mathcal{M}_{\mathrm{SM}}^{ij \rightarrow kl H}|^2\, .
\end{eqnarray*}\end{linenomath}
 
The best estimate and confidence intervals for $\tilde{d}$ in this analysis are determined by a fit of the predicted distribution of the \OOtext to that measured in data.
It has been shown in Ref.~\cite{HIGG-2015-06} that the \OOtext yields a significantly higher sensitivity in the determination of
$\tilde{d}$ than the CP-odd signed difference in the azimuthal angle $\Delta \phi_{jj}$ between the two tagging jets,
as suggested in Ref.~\cite{Hankele:2006ma}.
\section{ATLAS detector}
\newcommand{\AtlasCoordFootnote}{
ATLAS uses a right-handed coordinate system with its origin at the nominal interaction point (IP)
in the centre of the detector and the \(z\)-axis along the beam pipe.
The \(x\)-axis points from the IP to the centre of the LHC ring,
and the \(y\)-axis points upwards.
Cylindrical coordinates \((r,\phi)\) are used in the transverse plane,
\(\phi\) being the azimuthal angle around the \(z\)-axis.
The pseudorapidity is defined in terms of the polar angle \(\theta\) as \(\eta \equiv -\ln \tan(\theta/2)\).
Angular distance is measured in units of \(\Delta R \equiv \sqrt{(\Delta\eta)^{2} + (\Delta\phi)^{2}}\).}
 
The ATLAS experiment~\cite{PERF-2007-01,ATLAS-TDR-19,PIX-2018-001} at the LHC is a multipurpose particle detector
with a forward--backward symmetric cylindrical geometry and a near \(4\pi\) coverage in
solid angle.\footnote{\AtlasCoordFootnote}
It consists of an inner tracking detector surrounded by a thin superconducting solenoid
providing a \SI{2}{\tesla} axial magnetic field, electromagnetic and hadron calorimeters, and a muon spectrometer.
The inner tracking detector covers the pseudorapidity range \(|\eta| < 2.5\).
It consists of silicon pixel, silicon microstrip, and transition radiation tracking detectors.
Lead/liquid-argon (LAr) sampling calorimeters provide electromagnetic (EM) energy measurements
with high granularity.
A steel/scintillator-tile hadron calorimeter covers the central pseudorapidity range (\(|\eta| < 1.7\)).
The endcap and forward regions are instrumented with LAr calorimeters
for both the EM and hadronic energy measurements up to \(|\eta| = 4.9\).
The muon spectrometer surrounds the calorimeters and is based on
three large air-core toroidal superconducting magnets with eight coils each.
The field integral of the toroids ranges between \num{2.0} and \SI{6.0}{\tesla\metre}
across most of the detector.
The muon spectrometer includes a system of precision tracking chambers and fast detectors for triggering.
The integrated luminosity recorded by ATLAS is obtained with the LUCID-2 detector~\cite{LUCID2}.
 
A two-level trigger system is used to select events~\cite{TRIG-2016-01}.
The first-level trigger is implemented in hardware and uses a subset of the detector information
to reduce the accepted rate to at most \SI{100}{\kHz}.
This is followed by a software-based trigger that
reduces the accepted event rate to \SI{1}{\kHz} on average
depending on the data-taking conditions.
 
\section{Simulated event samples}
\label{sec:samples}
Samples of signal and background events were simulated using various Monte Carlo~(MC) event generators.
The generators and the PDF sets used for the hard-scattering process and the models used for the parton showers, hadronization, and underlying-event activity~(UEPS) are summarized in Table~\ref{tab:mcsamples}. In addition, the order of the total cross-section calculation is given.
 
Only Higgs boson production via VBF is considered as signal, including the signals observed as $\Htt$ decay and $\hwwlnln$ decay. The analysis is not sensitive to CPV in the $\HWW$ decay vertex and the shape of the \OOtext~is the same for the $\hwwlnln$ and $H\to\tau\tau \to \ell\ell 4\nu$ decay modes regardless of the value of $\dtilde$. The other Higgs boson production modes -- gluon--gluon fusion~(ggF $H$), $VH$, $t\bar{t}H$ -- are considered as background in this analysis, and all couplings other than the $HVV$ coupling were set to SM values.
All SM signal and background samples used in this analysis are the same as those employed in Ref.~\cite{HIGG-2017-07}, and the same normalization of those samples is used. The only exception is the normalization of the electroweak $Zjj$ process. Here, the leading-order (LO) cross section calculated by the \SHERPAV{2.2.1} generator~\cite{Gleisberg:2008ta,Gleisberg:2008fv,Cascioli:2011va,Hoeche:2009rj}
is multiplied by a factor of 1.7 to match the cross-section value measured by the ATLAS experiment at $\rts = \SI{13}\TeV$~\cite{STDM-2016-09}. An uncertainty of 25\% from the measured cross-section of the electroweak $Zjj$ process is applied to the normalization.
 
\begin{table}[h]
\centering
\caption{Overview of simulation tools used to generate signal and background processes and to model the UEPS. Details on the tunes used in the UEPS model
can be found in Ref.~\cite{HIGG-2017-07}. The PDF sets are also summarized. All Higgs boson events were
generated assuming $m_{H}= \SI{125}\GeV$. Alternative event generators and configurations used to estimate systematic uncertainties are shown in parentheses.
The prediction order in the last column refers to the cross section used to normalize the event sample.
}
\label{tab:mcsamples}
\resizebox{\textwidth}{!}{
\begin{tabular}{@{}l l l l l@{}}
\toprule
Process              & Matrix element                      & PDF set & UEPS model                           & Prediction order\\
& (alternative)                       &      & (alternative model)                &  for total cross section\\
\midrule
VBF $H$                 & \POWHEGBOXV{v2}~\cite{Nason:2004rx,Frixione:2007vw,Alioli:2010xd,Hamilton:2012np,Nason2010}           & PDF4LHC15 NLO~\cite{Butterworth:2015oua} & \PYTHIAV{8}~\cite{Sjostrand:2014zea} & approx. NNLO QCD + NLO EW~\cite{Ciccolini:2007jr,Ciccolini:2007ec,Bolzoni:2010xr}\\
&                                &                             & (\HerwigV{7}~\cite{Bahr:2008pv,Bellm:2015jjp}) & \\
ggF $H$                 & \POWHEGBOXV{v2}                                                  & PDF4LHC15 NNLO       & \PYTHIAV{8} & N$^{3}$LO QCD + NLO EW~\cite{Anastasiou:2015ema,Anastasiou:2016cez,Actis:2008ug,Anastasiou:2008tj}\\
& NNLOPS~\cite{Campbell:2012am,Hamilton:2013fea,Hamilton:2015nsa}     &                                                   &     (\HerwigV{7})      &  \\
$VH$                &  \POWHEGBOXV{v2}~\cite{Luisoni2013}            & PDF4LHC15 NLO                     & \PYTHIAV{8} & $qq/qg \rightarrow VH$: NNLO QCD + NLO EW~\cite{Brein:2003wg,Denner:2011id}\\
&   &   &   & $gg \rightarrow ZH$: NLO + NLL QCD~\cite{Altenkamp:2012sx,Harlander:2014wda} \\
$t\bar{t}H$                &  \MGkMCatNLOV{2.2.2}~\cite{Alwall:2014hca,Frederix:2012ps}            & NNPDF3.0LO~\cite{Ball:2014uwa}       & \PYTHIAV{8} & NLO QCD + NLO EW~\cite{ttH_NLO1,ttH_NLO2,ttH_NLO3, ttH_NLO4,ttH_NLO5,ttH_NLO6}\\
$W/Z$+jets    & \SHERPAV{2.2.1}~\cite{Hoeche:2012yf}                      & NNPDF3.0NNLO                     & \SHERPAV{2.2.1}~\cite{Schumann:2007mg}         & NNLO~\cite{Melnikov:2006kv,Anastasiou:2003ds} \\
& (\MGkMCatNLOV{2.2.2})&                  & (\PYTHIAV{8}) & \\
Electroweak $W/Zjj$                & \SHERPAV{2.2.1}                      & NNPDF3.0NNLO                      & \SHERPAV{2.2.1}        & LO \\
 
$VV/V\gamma^{\ast}$    & \SHERPAV{2.2.1}                         & NNPDF3.0NNLO                     & \SHERPAV{2.2.1}                  & NLO \\
$t\bar{t}$           & \POWHEGBOXV{v2}~\cite{Alioli:2011as}  & CT10~\cite{Lai:2010vv}       &\PYTHIAV{6.428}~\cite{Sjostrand:2006za} & NNLO+NNLL~\cite{Czakon:2013} \\
$Wt$                  & \POWHEGBOXV{v1}~\cite{Re:2010bp}       & CT10           &  \PYTHIAV{6.428} & NLO~\cite{Re:2010bp}                \\
\bottomrule
\end{tabular}
}
\end{table}
 
To simulate the presence of non-vanishing values of $\dtilde$ in the $HVV$ vertex, a matrix-element reweighting method is applied to the VBF SM signal sample. The weight is defined as the ratio of the squared ME value of the VBF process associated with a specific amount of CP mixing (given in terms of $\dtilde$) to that obtained from the SM. To extract the weights, the leading-order MEs from HAWK
are used for the $2 \rightarrow 2+H$ and $2 \rightarrow 3 + H$ processes separately. The MEs are evaluated using the four-momenta and particle identification codes of the initial- and final-state partons and the Higgs boson of each event. The reweighting procedure has been validated~\cite{HIGG-2015-06} against samples generated with \MGMCatNLO~\cite{Maltoni:2013sma} and proves to be a good approximation of a full NLO description of the process with non-vanishing values of $\dtilde$.
 
For all samples, a full simulation of the ATLAS detector response~\cite{SOFT-2010-01} using the $\GEANT4$ program~\cite{Agostinelli:2002hh} was performed.
The effect of multiple \pp interactions in the same and neighbouring bunch
crossings (pile-up) was included by overlaying minimum-bias events simulated
with \PYTHIAV{8} using the MSTW2008LO PDF~\cite{MSTW2008LO}  and the A2 set~\cite{ATL-PHYS-PUB-2012-003} of tuned parameters on each generated signal and background event.
The number of overlaid events was chosen such that the distribution of the average number of interactions per \pp bunch crossing in the simulation matches that observed in data.
 

\section{Event selection}
\label{sec:analysis}
In this analysis, events with at least two jets and a $\Htt$ decay candidate in the final state are selected. Decays of the $\tau$-leptons with all combinations of leptonic ($\tau \rightarrow \ell\nu\bar{\nu}$ with $\ell = e,\mu$) and hadronic ($\tau \rightarrow \mathrm{hadrons}\;\nu$) final states are considered.
In the following, the event preselection, which closely follows Ref.~\cite{HIGG-2017-07}, is summarized and the analysis strategy using gradient boosted decision trees (BDTs)~\cite{friedman2001} is described.
After data quality requirements~\cite{DAPR-2018-01}, the integrated luminosity of the $\rts = \SI{13}\TeV$ dataset used is $\SI{36.1}\ifb$.
The definition of the reconstructed objects as well as the triggers used in this analysis correspond to those used in Ref.~\cite{HIGG-2017-07}, where more details are given.
 
Depending on the reconstructed decay modes of the two $\tau$-leptons, events are separated into four analysis channels: the dileptonic same-flavour ($\tll$ SF), the dileptonic different flavour ($\tll$ DF), the semileptonic (\tlhad), and the fully hadronic (\thadhad) channel. All channels require an exact number of identified and isolated $\tau$-lepton decay candidates, i.e.\ electrons, muons, and visible products of hadronic $\tau$ decays (\tauhadvis), as defined in Ref.~\cite{HIGG-2017-07}, corresponding to their respective final state.
Events with additional $\tau$-lepton decay candidates are rejected. This ensures that the  selected data samples in the four channels do not overlap.
The two $\tau$-lepton decay candidates are required to be of opposite electric charge and to fulfil the requirements on the transverse momentum given in Table~\ref{tab:selection}.

\begin{table}
\caption{Summary of the event selection requirements for the four analysis channels.
In the case of the \pT requirements on the $\tau$-lepton decay candidates, the asterisk marks the lowest $\pT$ threshold, which varies depending on the trigger used.
Details of this are given in Ref.~\cite{HIGG-2017-07}.
The transverse momentum of the visible decay products of the $\tau$-lepton candidate with the higher (lower) transverse momentum is denoted by $\pTtl$ (\pTtsl).
The input variables used for the BDT training and the $\bdtscore$ threshold used to define the signal regions are also reported.}
\label{tab:selection}
\centering
\scalebox{0.90}{
\vspace{2mm}
\small
\begin{tabular}{c | c | c | c | c}
\toprule
Channel & $\tll$ SF                   & $\tll$ DF                       & \tlhad                               & \thadhad \\
\midrule
\multirow{8}{*}{ Preselection} & \multicolumn{4}{c}{Two isolated $\tau$-lepton decay candidates with opposite electric charge}                      \\
& $\pTtl > 19^{\ast}/15^{\ast}\,\GeV\, (\mu/e)$    &  $p_\text{T}^{e} > \SI{18}{\GeV}$    &     $\pTth > \SI{30}{\GeV}$    & $\pTtl > \SI{40}{\GeV}$         \\
&  $\pTtsl > 10/15^{\ast}\,\GeV\, (\mu/e)$   &  $p_\text{T}^{\mu} > \SI{14}{\GeV}$    &     $\pTtlep > 21^{\ast}\,\GeV$       & $\pTtsl > \SI{30}{\GeV}$         \\
& \multicolumn{2}{c|}{$\mcoll > m_{\PZ}-\SI{25}{\GeV}$}     & $\mT < \SI{70}{\GeV}$                & $0.8 < \dRtt < 2.5$         \\
& $30<\mll<\SI{75}{\GeV}$    & $30<\mll<\SI{100}{\GeV}$     &                                      & $|\detatt| < 1.5$         \\
& $\met > \SI{55}{\GeV}$     & $\met > \SI{20}{\GeV}$       &                                      & $\met > \SI{20}{\GeV}$ \\
& $\METHPTO > \SI{55}{\GeV}$ &                              &                                      &          \\
& \multicolumn{3}{c|}{$N_\text{\Pqb-jets} = 0$}                          &          \\
\midrule
\multirow{2}{*}{VBF topology} & \multicolumn{4}{c}{$N_\text{jets} \geq 2$, $\pTjsl > \SI{30}{\GeV}$, $\mjj > \SI{300}{\GeV}$, $|\detajj| > 3$} \\
 
& \multicolumn{3}{c|}{$\pTjl > \SI{40}{\GeV}$}                                                      & $\pTjl > \SI{70}{\GeV}, |\eta_{j_1}| < 3.2$ \\
\midrule
\multirow{3}{*}{BDT input variables} & \multicolumn{4}{c}{\MMC, \mjj, \dRtt, \Cjjtl, \Cjjtsl, \pTtot} \\
& \multicolumn{2}{c|}{\mttvis, $\mT^{\tau_1,\met}$, \pTjssl}  & \multicolumn{2}{c}{\CMET}  \\
&    \dphitt     &   $\met/\pT^{\tau_1}$, $\met/\pT^{\tau_2}$ & \mttvis, $|\detatt|$ & \pTttMET, $|\detatt|$  \\
\midrule
Signal region & \multicolumn{2}{c|}{$\bdtscore > 0.78$} & $\bdtscore > 0.86$ & $\bdtscore > 0.87$ \\
\bottomrule
\end{tabular}}
\end{table}
 
The event selection for the four analysis channels is summarized in Table~\ref{tab:selection}.
In the $\tll$ and $\thadhad$ channels, only events with missing transverse momentum $\met > \SI{20}{\GeV}$ are selected to reject events without neutrino candidates.
To suppress the large background from \Zll production in the $\tll$ SF channel, the requirement on $\met$ is tightened. Furthermore, an additional requirement is imposed on the quantity \METHPTO, obtained from an \met calculation considering only contributions from reconstructed objects and neglecting contributions from inner-detector tracks originating from the vertex of the hard-scattering process, but not associated with any of the reconstructed objects.
In addition, a requirement on the invariant mass of the two light leptons, $\mll$, is applied in the $\tll$ channels.
A requirement on the di-\Pgt mass calculated in the collinear approximation~\cite{Ellis:1987xu} of $\mcoll > m_{\PZ}-\SI{25}{\GeV}$ is introduced in the $\tll$ channels
to ensure orthogonality between this analysis and the analysis of $\HWW\to\Pl\Pgn\Pl\Pgn$~\cite{HIGG-2013-13}, which has a similar final state.
In the \tll and \tlhad channels, the top quark background is suppressed by requiring that no jet with $\pT>\SI{25}{\GeV}$ and $|\eta| < 2.5$ contains $\Pqb$-hadrons ($\Pqb$-jets). A multivariate algorithm~\cite{PERF-2012-04, ATL-PHYS-PUB-2016-012} is used to identify and select $\Pqb$-jets with a working point corresponding to an average efficiency of 85\%,
as measured on a sample from top quark pair production.
Low transverse mass\footnote{The transverse mass is defined as $m_{\mathrm{T}}=\sqrt{2p_{\mathrm{T}}^{\ell} \
E_{\mathrm{T}}^{\mathrm{miss}} \cdot (1-\cos{\Delta{\phi}})}$, where
$\Delta{\phi}$ is the azimuthal separation between the directions of the
lepton and the missing transverse momentum.} ($\mT < \SI{70}{\GeV}$)
is required in the \tlhad channel to reject events with leptonic \PW decays.
Requirements on the angular distance between the visible products of the two selected \Pgt{}-lepton decays, \dRtt, and their
pseudorapidity difference, $|\detatt|$, are applied in the $\thadhad$ channel to reject non-resonant background events.
 
To select Higgs boson events produced by VBF, all channels require at least two jets with transverse momentum of the leading jet $\pTjl > \SI{40}{\GeV}$ and of the subleading jet $\pTjsl > \SI{30}{\GeV}$, a large invariant mass of the two leading jets, $\mjj > \SI{300}{\GeV}$, and a pseudorapidity separation of $|\detajj| > 3$.
In the $\thadhad$ channel, the requirements on the leading jet are raised to $\pTjl > \SI{70}{\GeV}$ and $|\eta_{j_1}| < 3.2$ to achieve a uniform trigger selection efficiency as a function of $\pTjl$.
This selection is referred to as the VBF event selection in the following.
 
To construct a region enriched in VBF signal events, BDTs trained to discriminate between the VBF signal and the backgrounds are used in all channels. Kinematic variables used in the BDT training can be categorized as follows:
\begin{itemize}
\item Properties of the Higgs boson which discriminate against all background processes without a Higgs boson: the visible mass of the di-\Pgt system, $\mttvis$, the transverse momentum of the $\tau\tau E_{\text{T}}^{\text{miss}}$ system, \pTttMET, and the reconstructed Higgs boson mass, \MMC, determined using the missing-mass calculator (MMC)~\cite{MMC}.
 
\item Properties of a resonant di-\Pgt decay which discriminate against processes with jets that are misidentified as $\tau$-decay candidates (referred to as ``Misidentified $\tau$''):
the angular distance \dRtt, the difference in pseudorapidity $|\detatt|$, and the difference in azimuth \dphitt between the two visible \Pgt-leptons.
In addition, the transverse momentum ratio $\met/\pT^{\tau_1}$ ($\met/\pT^{\tau_2}$) between the $\met$ and the leading (subleading) \Pgt-candidate as well as the transverse mass of the $\met$ and the leading \Pgt-candidate, $\mT^{\tau_1,\met}$, is used.
Furthermore, the azimuthal centrality of \MET, \CMET, which quantifies the angular direction of the missing transverse momentum relative to the
visible $\tau$-decay products in the transverse plane, is constructed.\footnote{
$C(\phi^\text{miss})$ is defined as $(A+B)/\sqrt{A^2 + B^2}$, where $A = \sin(\phi_{\MET} - \phi_{\tau_2})/\sin(\phi_{\tau_1} - \phi_{\tau_2})$ and $B = \sin(\phi_{\tau_1} - \phi_{\MET})/\sin(\phi_{\tau_1} - \phi_{\tau_2})$. }

\item Properties of the VBF topology:  \mjj, the total transverse momentum \pTtot, which is defined as the transverse momentum of the system composed of all objects
in a VBF event ($\tau_1$, $\tau_2$, $j_1$, $j_2$, \met), $\eta$-centralities, \Cjjtl and \Cjjtsl, of each \Pgt-candidate relative to the pseudorapidity
of the two leading jets,\footnote{
$C_{jj}(\tau) = \exp \left[ \frac{-4}{(\eta_{j_1} - \eta_{j_2})^2}\left(\eta_{\tau} - \frac{\eta_{j_1} + \eta_{j_2}}{2}\right)^2\right]$ ,
where $\eta_{\tau}$, $\eta_{j_1}$ and $\eta_{j_2}$ are the pseudorapidities of the \Pgt-candidate and the two leading jets, respectively.  This variable
has a value of unity when the object is halfway in $\eta$ between the two jets, $1/$e when the object is aligned with one of the jets, and $<1/$e when the
object is not between the jets in $\eta$.} and the transverse momentum of the third leading jet $\pTjssl$ which is set to zero for events with exactly two
jets.
 
\end{itemize}
 
\begin{figure}[h!]
\centering
\subfloat[]{\includegraphics[width=0.43\linewidth]{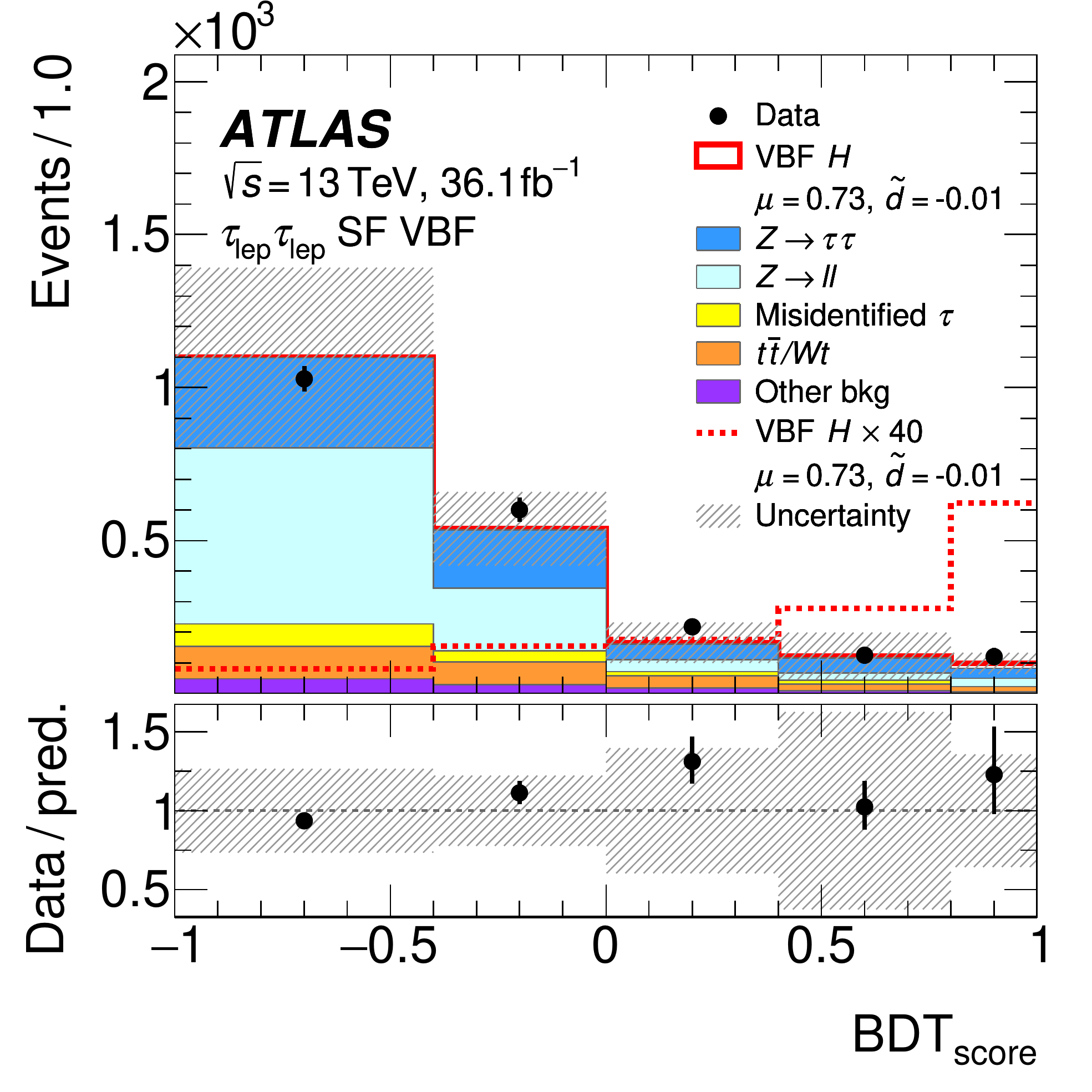} }
\subfloat[]{\includegraphics[width=0.43\linewidth]{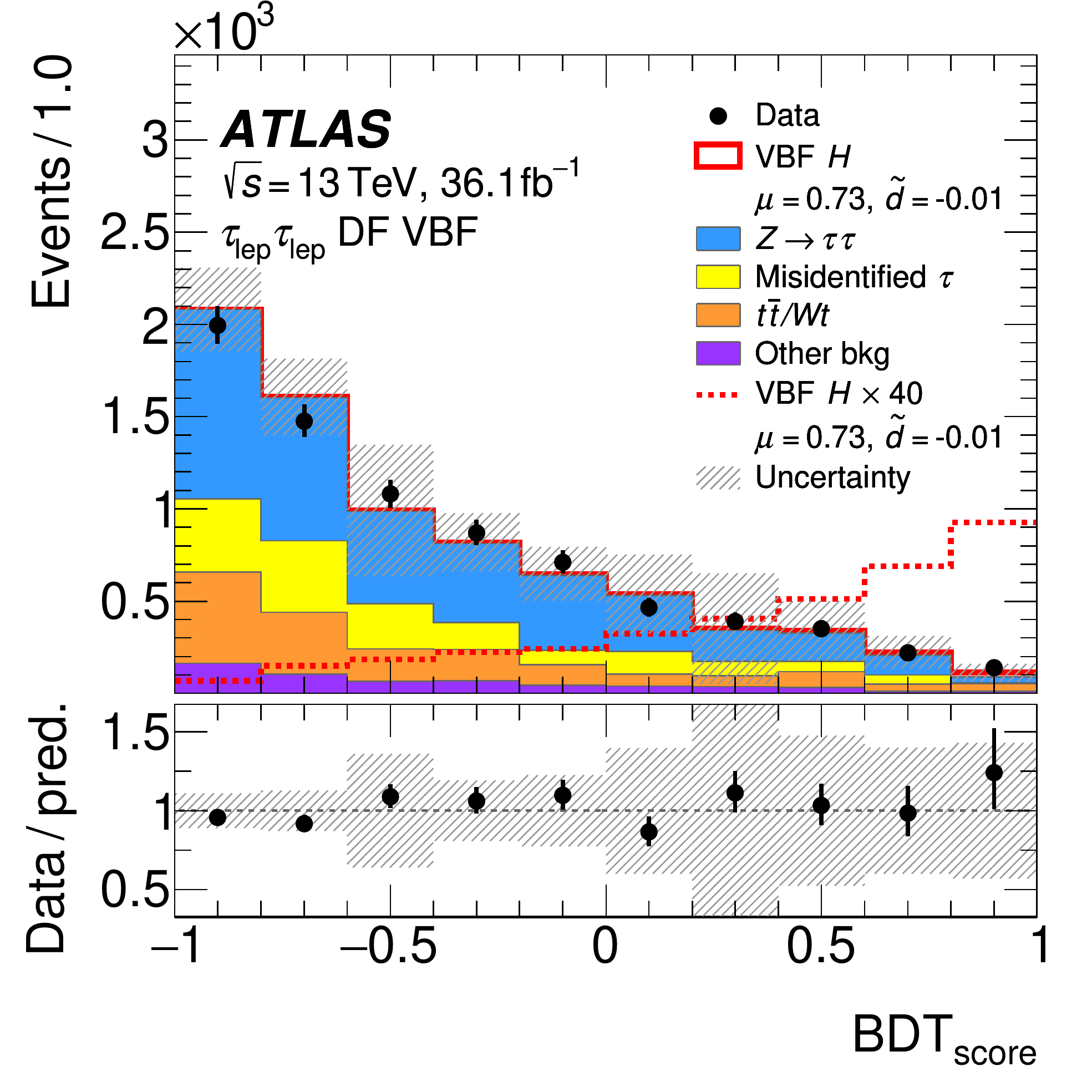} }\\
\subfloat[]{\includegraphics[width=0.43\linewidth]{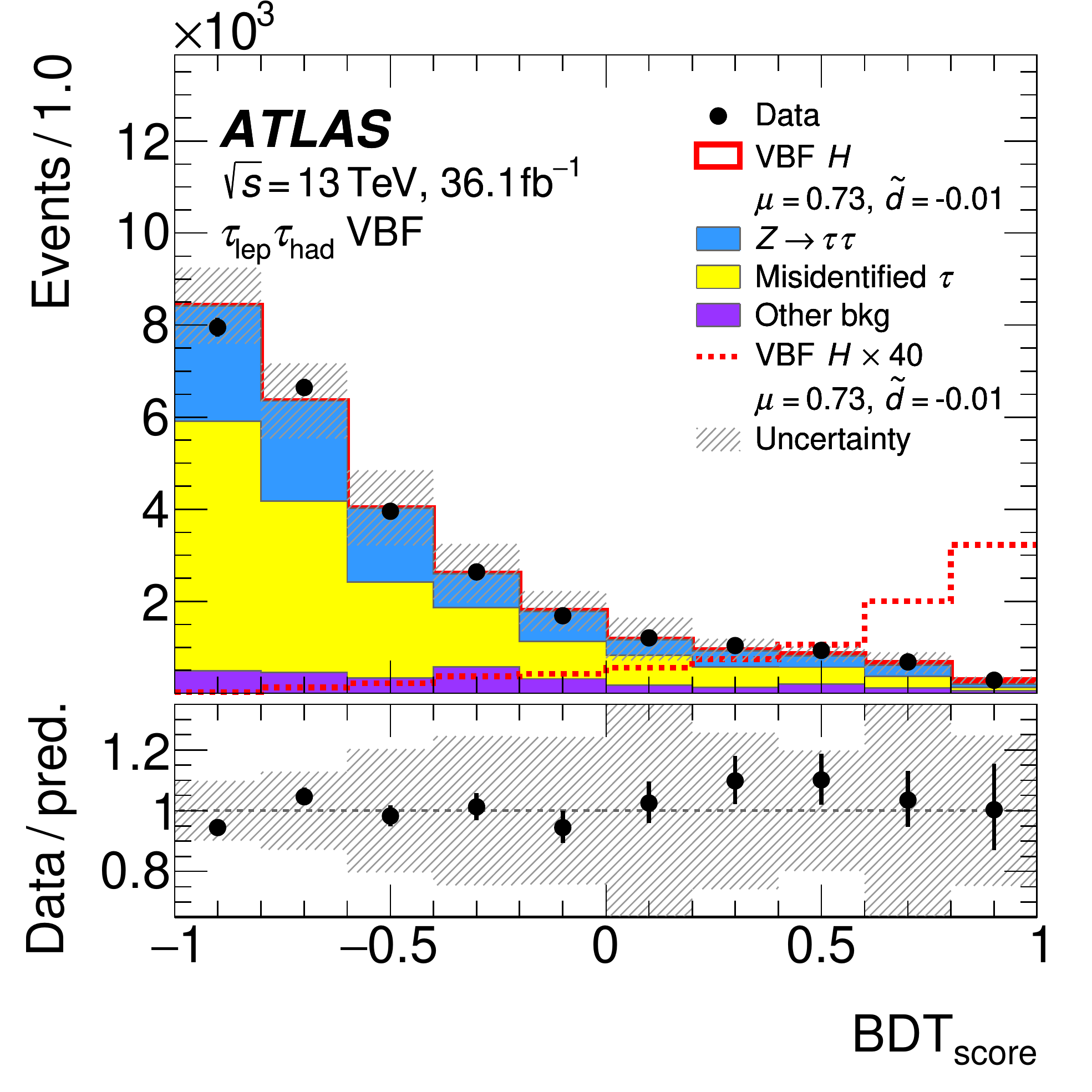} }
\subfloat[]{\includegraphics[width=0.43\linewidth]{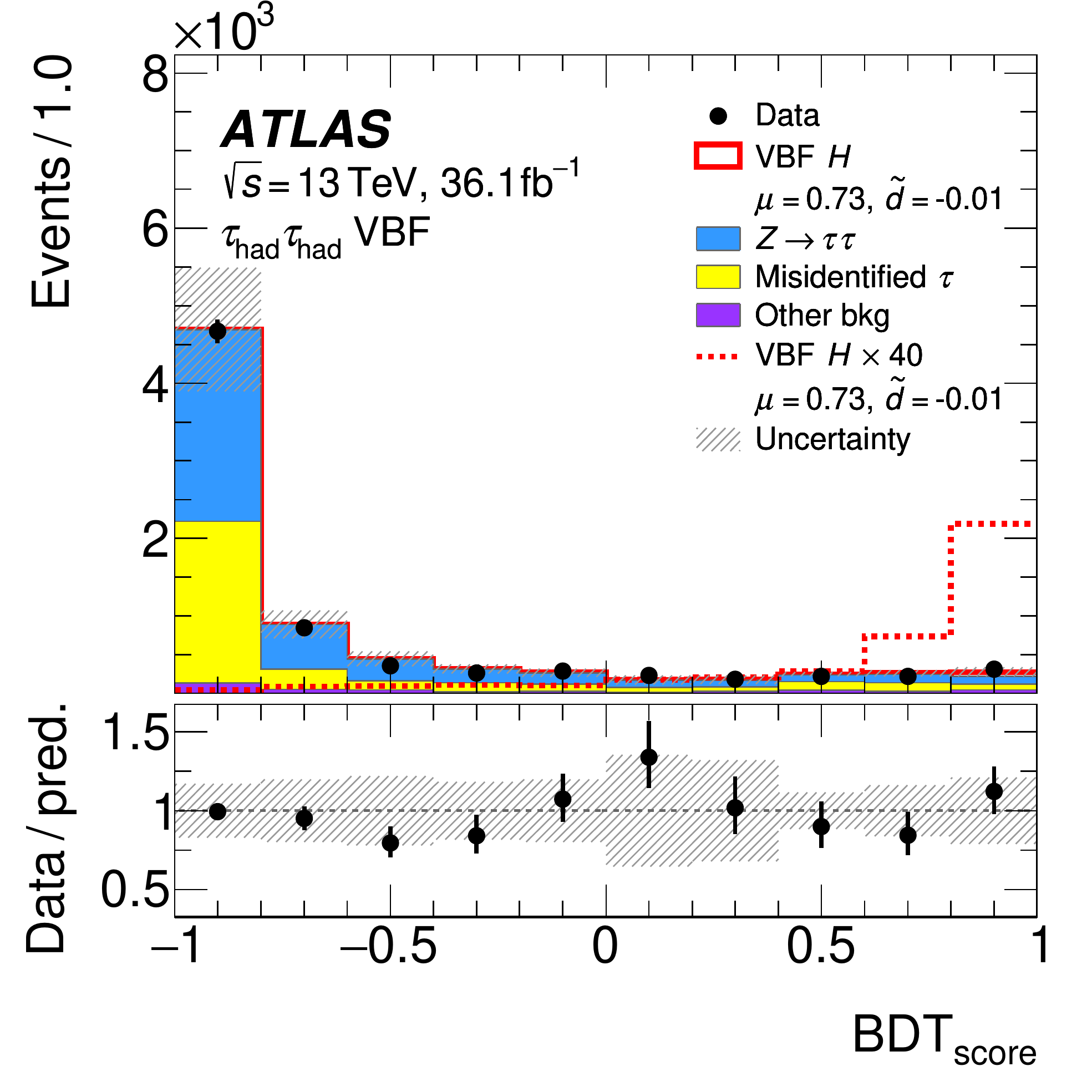} }
\caption{Post-fit $\bdtscore$ distributions after the VBF event selection for the (a) \tll~SF, (b) \tll~DF , (c) \tlhad~and (d) \thadhad analysis channels. The ratios of the data to the prediction are shown in the lower panels. The observed VBF signal ($\mu = 0.73$, $\dtilde = -0.01$) is shown with the solid red line on the top of the histogram stack. ``Other bkg'' denotes all background contributions not listed explicitly in the legend. The dashed line shows the observed VBF signal scaled up by a factor of 40 and is not part of the histogram stack. The size of the combined statistical, experimental, and theoretical uncertainties in the background is indicated by the hatched bands.
\label{fig:bdts}}
\end{figure}
 
The most important variables in the training are \MMC, \mjj, and \Cjjtl.
The resulting BDT score ($\bdtscore$) distributions are shown in Figure~\ref{fig:bdts} for events surviving the VBF event selection and show the ability of the BDT to separate the signal process from background processes. All figures in this Letter use signal strength $\mu$ (defined as the
ratio of the measured cross section times branching ratio to the SM prediction for the VBF signal process), background normalizations, and systematic uncertainties as fitted by the final statistical analysis discussed
in Section~\ref{sec:fitmodel} and referred to as \textit{post-fit}. The signal purity increases at high values of $\bdtscore$. A threshold value of $\bdtscore$ is used to define the final signal region~(SR) in each channel. This threshold is chosen to yield a high signal significance and is given in Table~\ref{tab:selection} for each channel.
The efficiency of the signal selection relative to the VBF event selection is 32\% (27\%) for the \tll~SF (\tll~DF) channel, 29\% for the \tlhad channel, and 49\% for the \thadhad channel. The efficiency for the sum of background processes, on the other hand, is 1.5\% (0.8\%) for the \tll~SF (\tll~DF) channel, 0.4\% for the \tlhad channel, and 1.1\% for the \thadhad channel.
In each SR the \OOtext is then used to probe for CPV. No dependence of the mean values of the \OOtext on $\bdtscore$ is observed, confirming that the SR selection criteria
do not introduce a CP asymmetry.

 
\section{Background estimation}
Several background processes contribute to the SR event yields in the four analysis channels. The dominant contributions in the \tll~DF, \tlhad, and \thadhad channels arise from \Ztt production and from light- and heavy-flavour jets that are misidentified as prompt leptonic or hadronic $\tau$ decays. The misidentified $\tau$ decays in the \tll and \tlhad channels originate largely from $W$+jets production with smaller contributions from multijet and top quark production, while in the \thadhad channel the contribution from multijet production dominates. In the \tll~SF channel the contribution from \Zll production is dominant. Other background contributions in all analysis channels originate from top quark pair and associated $Wt$ production (denoted by ``$t\bar{t} / Wt$'' in the following), diboson production, and other Higgs boson production modes.
 
Background contributions with prompt leptonic or hadronic $\tau$ decays are estimated from simulation, while the estimation of the background contribution from misidentified $\tau$ decays is mostly data-driven~\cite{HIGG-2017-07}.
Dedicated control regions~(CR) are defined in data to normalize the predictions of the following background processes: $\Ztt$ (for all channels), $t\bar{t} / Wt$ and $\Zll$ (only for the $\tll$ channels), and the misidentified $\tau$ decays (only for the $\thadhad$ channel).
All other background processes with prompt $\tau$ decays (including other Higgs boson production modes) are normalized to their SM prediction.

To construct a CR for $\Ztt$ production, the SR requirement on the \bdtscore (given in Table~\ref{tab:selection}) is inverted for each analysis channel. This CR is called the ``low-\bdtscore CR'' in the following. Since the purity of $\Ztt$ production in the low-\bdtscore CR ranges from 30\% to 54\% depending on the analysis channel, $\Ztt$ production is normalized to data in the $Z$ boson mass peak of the \MMC distributions, shown in Figure~\ref{fig:MMC}.
In the fit the \Ztt normalization is correlated across all analysis channels and the fit yields a normalization factor of $0.93\pm0.08$.
To ensure that the normalization is valid in the SR, the modelling of the $Z$-boson and jet kinematic properties was checked in a validation region which is composed of $\Zll$ events with kinematic properties similar to those of the $\Ztt$ events in the VBF region of each analysis channel. This region is defined by selecting two same-flavour leptons of opposite charge with a dilepton mass of $\mll > \SI{80}{\GeV}$ and low missing transverse momentum ($\met < \SI{55}{\GeV}$). All VBF selection requirements given in Table~\ref{tab:selection} are applied as well.
As in Ref.~\cite{HIGG-2017-07}, a slight positive slope
in the ratio of the data to the \SHERPA simulation as a function of \mjj is observed.
In this analysis, the simulated \Ztt and \Zll events are reweighted to the observed \mjj distribution after the VBF event selection, which results in a small change in the acceptance of \Ztt and \Zll events in the SR.
 
\begin{figure}[h]
\centering
\subfloat[]{\includegraphics[width=0.43\linewidth]{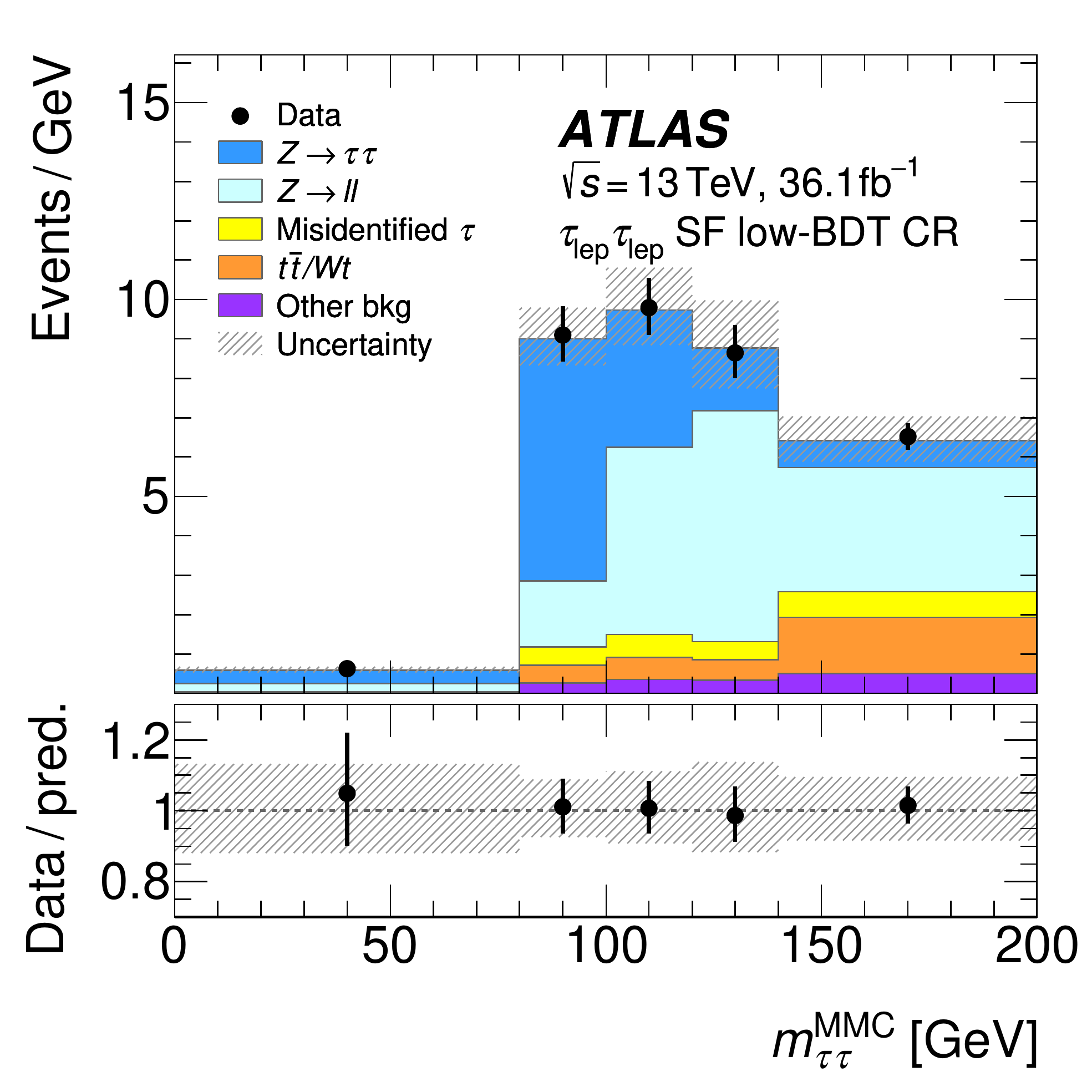} }
\subfloat[]{\includegraphics[width=0.43\linewidth]{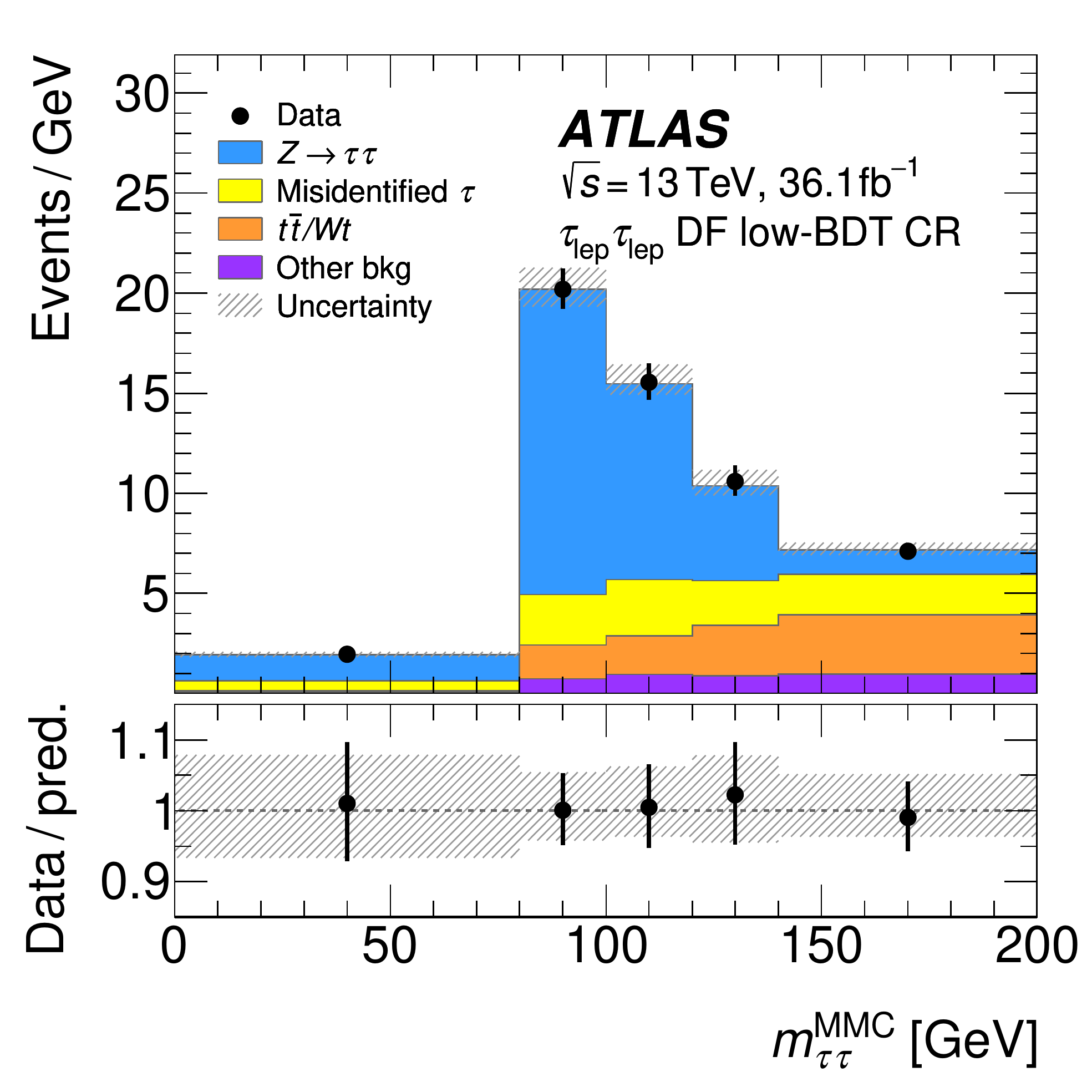} }\\
\subfloat[]{\includegraphics[width=0.43\linewidth]{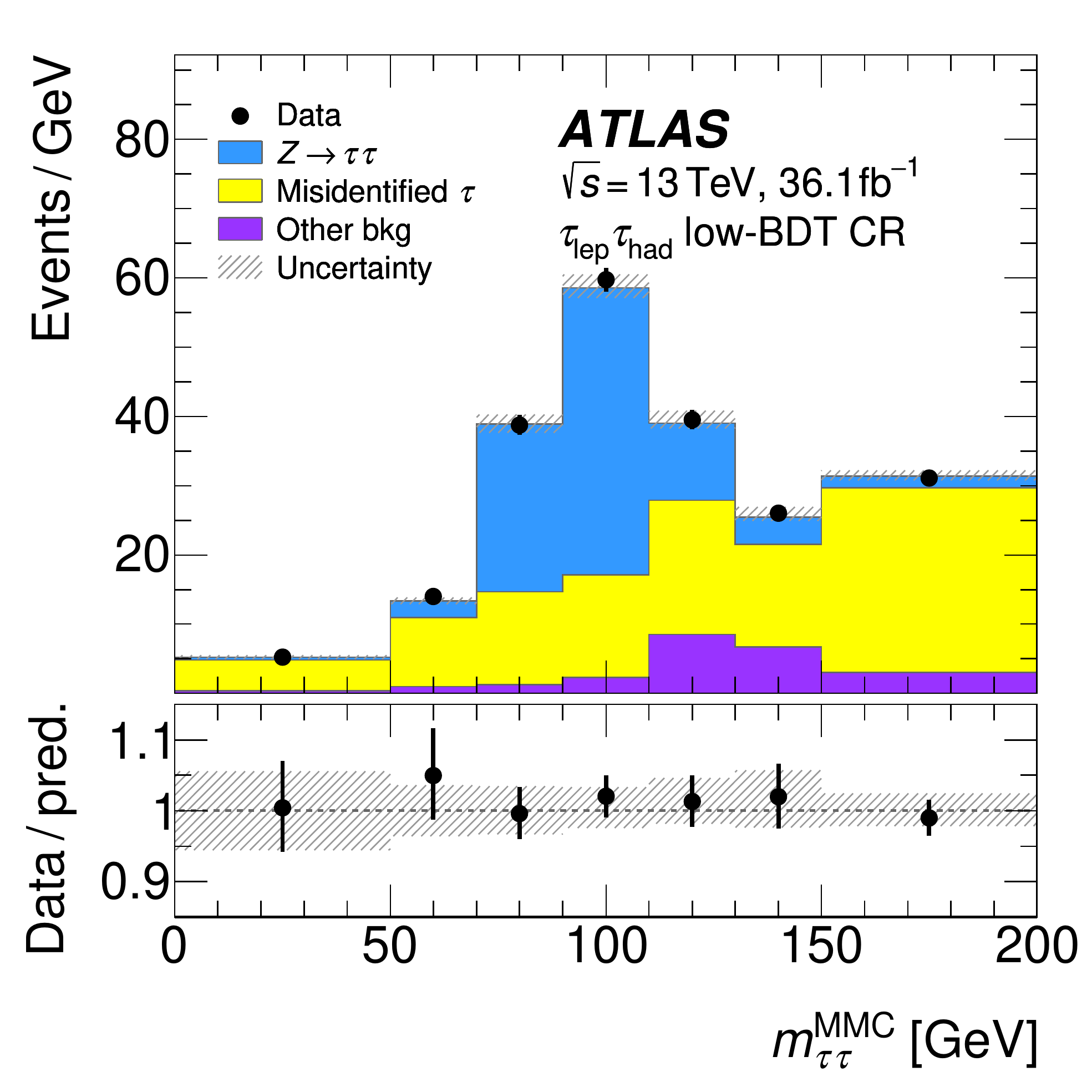} }
\subfloat[]{\includegraphics[width=0.43\linewidth]{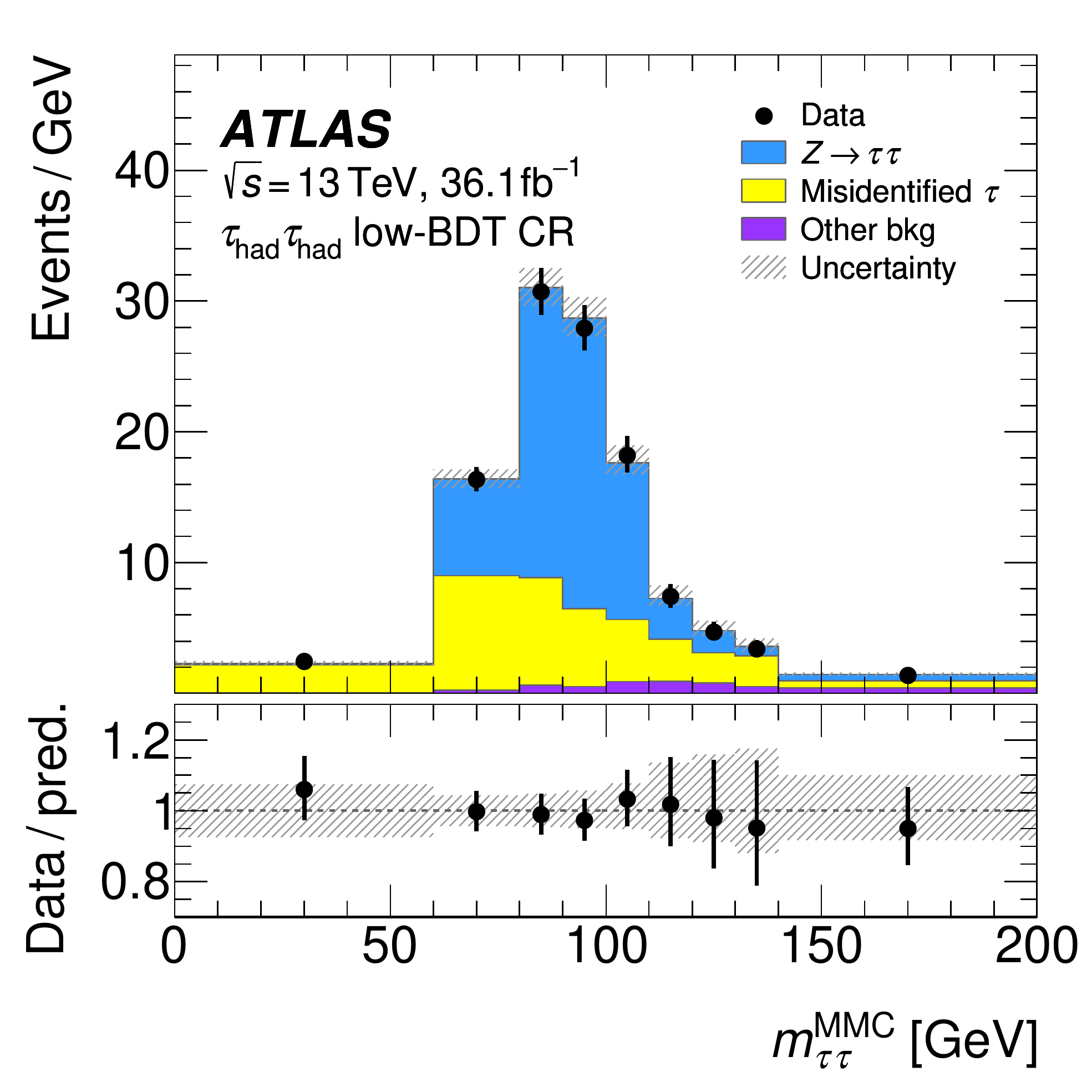} }
\caption{Post-fit \MMC distributions in the low-\bdtscore CR for the (a) \tll~SF, (b) \tll~DF , (c) \tlhad~and (d) \thadhad analysis channels. The ratios of the data to the prediction are shown in the lower panels. The contamination of the CR by signal is negligible.
``Other bkg'' denotes all background contributions not listed explicitly in the legend. The size of the combined statistical, experimental, and theoretical uncertainties in the background is indicated by the hatched bands. The rightmost bins in each of the subfigures include event yields with \MMC values larger than the shown range.
\label{fig:MMC}}
\end{figure}
 
In each of the two \tll channels, a top quark CR is defined by inverting the veto on $b$-tagged jets and not applying the selection on the \bdtscore. The normalization of $t\bar{t}/Wt$ production is constrained by the event yield in these CRs, corresponding to a normalization of $1.16\pm0.06$ from the combined fit to the data.
Additionally, another CR is defined to normalize the $\Zll$ process for the \tll~SF channel. Again, the selection on the \bdtscore is not applied, and the requirement on the dilepton invariant mass is changed to $80<\mll<\SI{100}{\GeV}$. The observed event yield in the $\Zll$ CR constrains the normalization of simulated \Zll events in the \tll~SF channel to $1.0 \pm 0.4$.
 
In the \thadhad channel, the background from misidentified hadronic $\tau$ decays is dominated by multijet events. This background process is modelled using a template extracted from $\tauhadvis$ candidates with one, two, or three associated tracks that pass all selection requirements, but fail the opposite-charge requirement. Before the final fit, the template is normalized to data by a fit of the $|\detatt|$ distribution after the preselection, but removing the requirement on $|\detatt|$. In the final fit the template is normalized to data in the \MMC distribution of the low-\bdtscore CR in the \thadhad channel. Then, the multijet background is normalized with a factor of $0.99\pm0.09$ relative to the pre-fit normalization.
 
The modelling of the \OOtext distribution for the background processes is validated in all CRs. Figure~\ref{fig:OO_CRs} shows \OOtext distributions in the low-\bdtscore CR for all analysis channels, where the background processes have been normalized to the result of the fit. Neither the observed nor the predicted distributions in any CR show hints of an asymmetry or non-vanishing mean values of the \OOtext caused by event reconstruction and selection within uncertainties. The data and the predicted distributions are observed to
be compatible within uncertainties here as well as in the top quark and $\Zll$ CRs of the \tll channels.
 
\begin{figure}[h]
\centering
\subfloat[]{\includegraphics[width=0.43\linewidth]{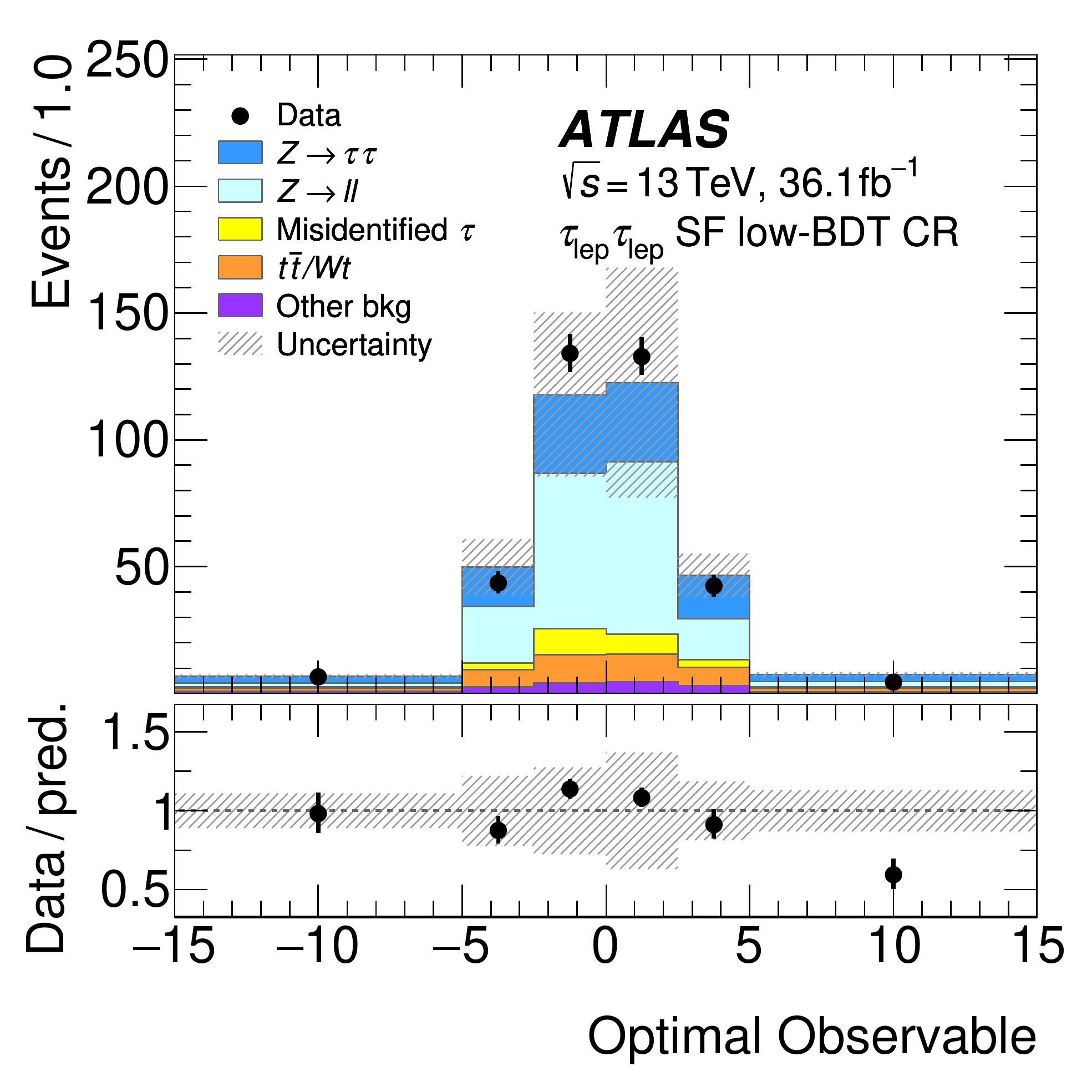} }
\subfloat[]{\includegraphics[width=0.43\linewidth]{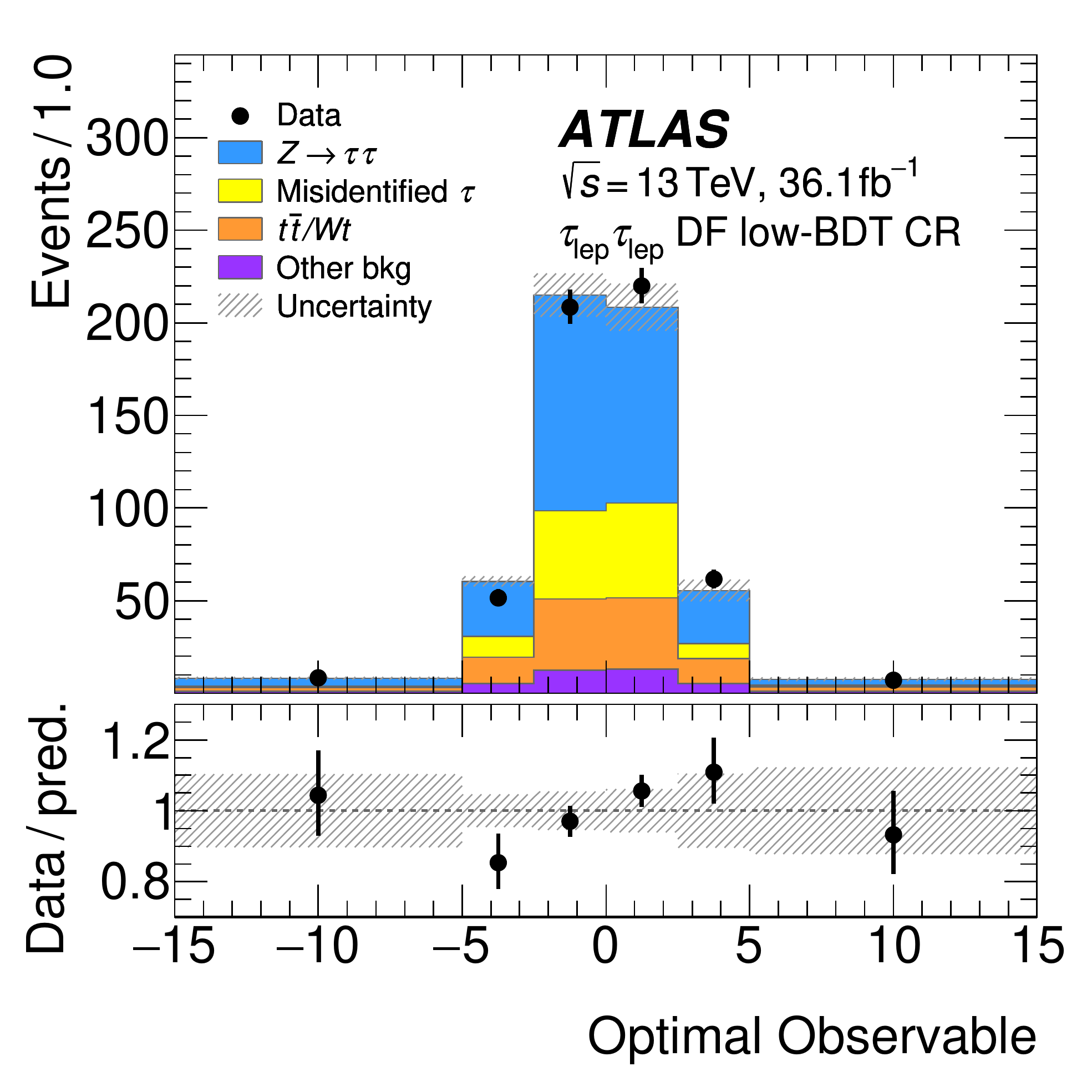} }\\
\subfloat[]{\includegraphics[width=0.43\linewidth]{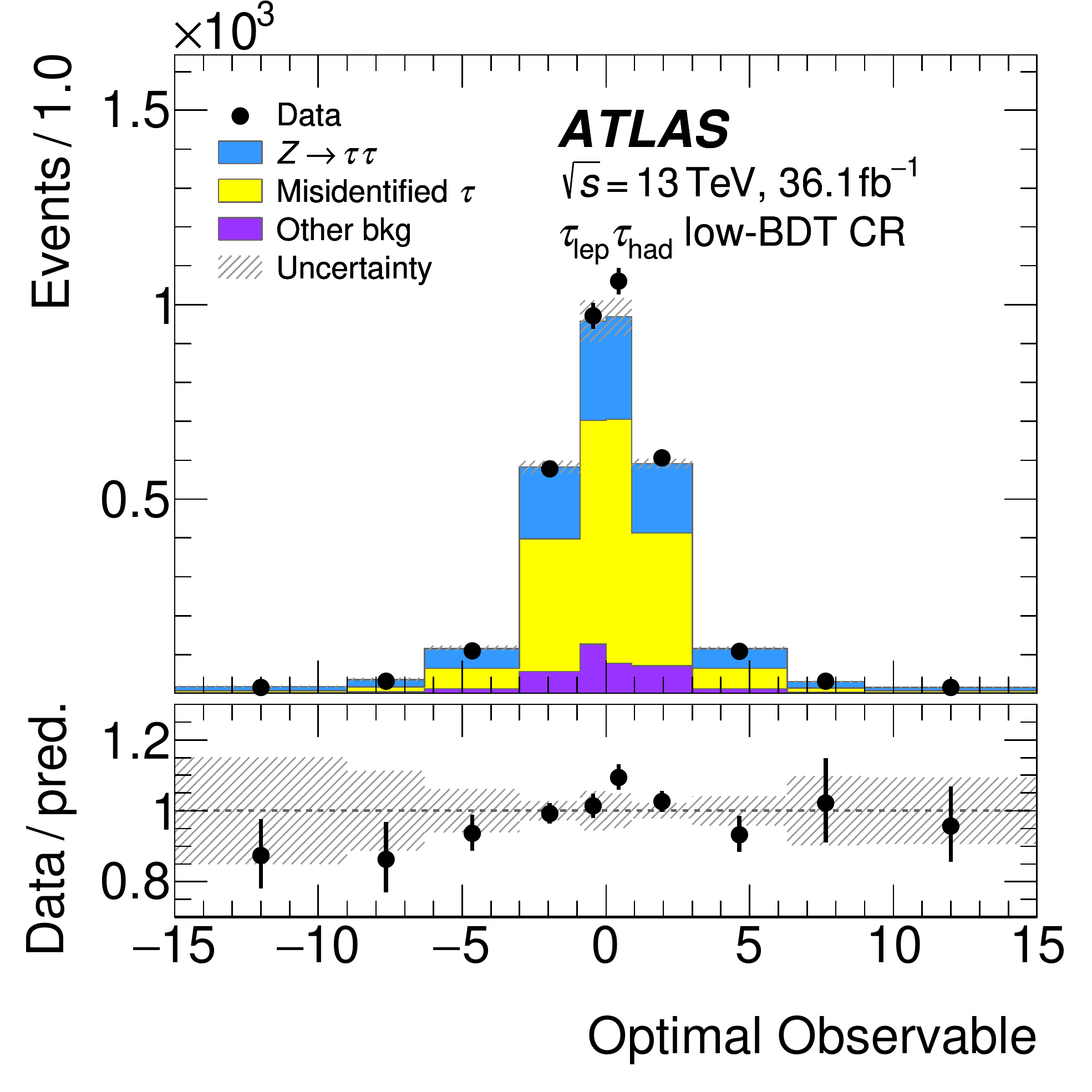} }
\subfloat[]{\includegraphics[width=0.43\linewidth]{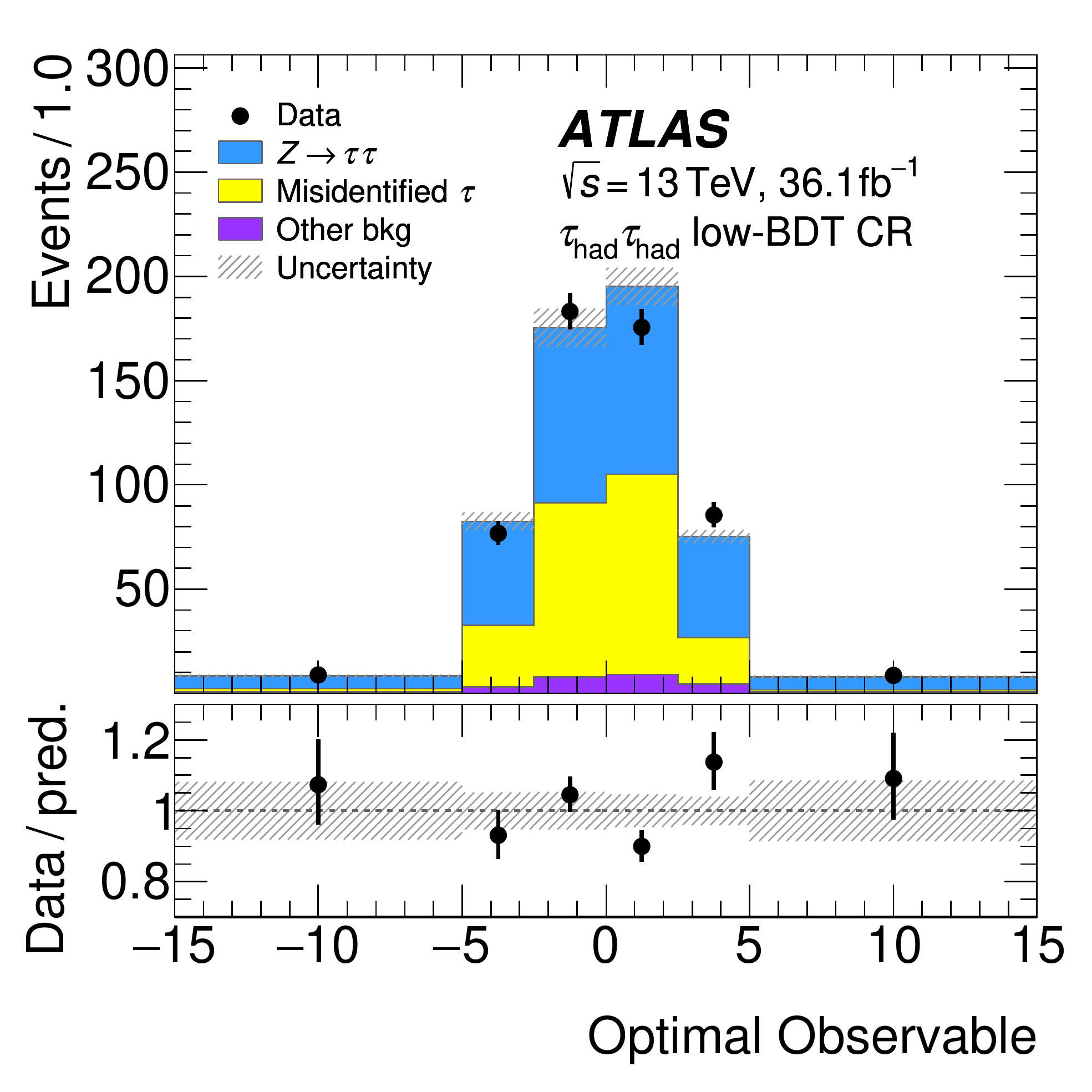} }
\caption{Post-fit \OOtext distributions in the low-\bdtscore CR for the (a) \tll~SF, (b) \tll~DF , (c) \tlhad~and (d) \thadhad analysis channels. The ratios of the data to the prediction are shown in the lower panels. The contamination of the CR by signal is negligible.
``Other bkg'' denotes all background contributions not listed explicitly in the legend. The size of the combined statistical, experimental, and theoretical uncertainties in the background is indicated by the hatched bands.
\label{fig:OO_CRs}}
\end{figure}


\section{Systematic uncertainties}
\label{sec:systematics}
The effects of the systematic uncertainties on the yields in both the SRs and CRs and on the shape of the \OOtext in the SRs, as well as the \MMC distributions in the CRs, are evaluated following the procedures in Ref.~\cite{HIGG-2017-07}. No sources of systematic uncertainties introduce a significant asymmetry in the \OOtext distribution.
The sources of uncertainty can be grouped into two categories: experimental and theoretical. The dominant experimental uncertainties stem from the determination of the jet energy resolution and scale~\cite{PERF-2016-04}, the $\tauhadvis$ energy scale and resolution~\cite{ATL-PHYS-PUB-2015-045}, and the $\tauhadvis$ reconstruction and identification efficiencies~\cite{ATLAS-CONF-2017-029}. Other sources of uncertainty are the electron (muon) energy (momentum) scale and resolution, lepton identification and isolation~\cite{PERF-2017-03,PERF-2017-01,PERF-2015-10}, missing transverse momentum reconstruction~\cite{PERF-2016-07}, $b$-tagging efficiency~\cite{PERF-2012-04,ATLAS-CONF-2018-001}, modelling of pile-up, and luminosity measurement~\cite{ATLAS-CONF-2019-021}. The luminosity uncertainty of 2.1\%~\cite{ATLAS-CONF-2019-021} is only applied to the VBF signal and to background processes normalized to theoretical predictions.
Uncertainties in backgrounds from misidentified $\tau$-leptons arise from the limited statistical precision of the data-driven templates and corrections used, from closure tests performed in regions where the $\tau$-leptons are required to have the same charge, and from the subtraction of the electroweak contributions.
 
Theoretical uncertainties affecting the total cross section are evaluated for the Higgs boson production cross sections for ggF~$H$, $VH$, and $t\bar{t}H$ production by varying the QCD factorization and renormalization scales as well as the PDF model following the recommendations in Ref.~\cite{deFlorian:2016spz}. Also, uncertainties in the $\Htt$ and $\HWW$ branching ratios are considered~\cite{deFlorian:2016spz}.
Theoretical uncertainties in the MC modelling are considered for the VBF and gluon--gluon fusion production of the Higgs boson as well as for $Z\to \tau \tau$ production. For all simulated background contributions other than $Z\to \tau \tau$, no theoretical uncertainties are considered, as their impact is negligible. Uncertainties in MC modelling are assessed by a comparison between nominal and alternative event generators and UEPS models, as indicated in Table~\ref{tab:mcsamples}. In addition, the effects of QCD factorization and renormalization scale variations, matching-scale variations (in the case of $Z\to \tau \tau$ only), and PDF model uncertainties are evaluated.
As an additional uncertainty in the $Z\to \tau \tau$ and \Zll processes, the full difference between the sample reweighted to the observed \mjj distribution and the sample without reweighting is applied to the full analysis.
An uncertainty to account for the signal reweighting procedure described in Section~\ref{sec:samples} was considered and found to be negligible. The uncertainty due to limited MC sample size is evaluated for the sum of all MC-based background processes in each analysis bin.
 
\section{Fitting procedure}
\label{sec:fitmodel}
 
The estimate of \dtilde~is obtained using a binned maximum-likelihood fit (ML-fit) performed simultaneously
on the SRs and all introduced CRs, which are
included in order to constrain background normalizations and nuisance parameters describing the
systematic uncertainties.
The ML-fit uses the distribution of the \OOtext~in each of the four high-\bdtscore~SRs,
one for each analysis channel. The \MMC distributions in the
low-\bdtscore~region for each channel are included in the ML-fit, and so are the event
yields in the $Z\rightarrow\ell\ell$ (\tll SF) and top quark (\tll SF and DF) CRs.
 
The inclusion of the \MMC distributions in the low-\bdtscore~regions provides the main constraint
on the \Ztt~normalization, which is free to float in the ML-fit. The $Z\rightarrow\ell\ell$ background
in the \tll~SF channel and top quark backgrounds in the \tll~SF and DF channels are also
free to float, and their contributions are constrained by the inclusion of CRs
in the ML-fit.
 
The normalization of the signal
is not constrained in the ML-fit, so that the analysis only exploits the shape of the distribution of the
\OOtext~in the estimation of \dtilde.  Any possible model-dependence of the cross section
on CP-mixing scenarios is not exploited. The relative contribution of the two
Higgs boson decay modes ($\Htt$ and $\HWW$) to the signal (relevant only for the \tll~channel) is
assumed to be correctly predicted by the SM.
All other Higgs boson production modes for these decays are considered as background and are normalized to their
SM predicted yields.
 
The ML-fit uses a binned likelihood function $\mathcal{L}(\boldsymbol{x}; \mu, \boldsymbol{\theta})$,
which is a function of the data $\boldsymbol{x}$, the free-floating signal strength $\mu$,
and nuisance parameters $\boldsymbol{\theta}$ corresponding to the systematic uncertainties
mentioned in Section~\ref{sec:systematics}. The likelihood function is evaluated for each
\dtilde~hypothesis using the relevant reweighted signal templates defined in Section~\ref{sec:samples},
with the background model unchanged, and a negative log-likelihood (NLL) curve can then be constructed
as a function of \dtilde.
 
The parameter of interest, \dtilde, is obtained at the point
where the NLL curve reaches a minimum. Central confidence intervals are obtained
by reading off the points on the NLL curve which exceed the minimum value by a certain amount.
 
 
\section{Results}
 
For a CP-even Higgs boson, the mean value of the \OOtext~for the
signal and background processes is expected to be zero if
any effects from the rescattering of new particles in loops can be
neglected. However, CP-violating effects could result in
the mean value of the \OOtext~in data deviating from zero, allowing an almost
model-independent test for CP-violating effects in this measurement.
 
The observed values for the mean of the \OOtext~in data, along with their statistical
uncertainties, are summarized in Table~\ref{tab:mean_value_OO} for the four channels in this
analysis, as well as their combination. The combined mean is obtained by weighting the
mean value of each individual channel by the inverse of its respective variance.
These values are fully consistent with zero, so no evidence of CPV is observed.
 
\begin{table}[h!]
\centering
\caption{Mean values of the \OOtext~with statistical uncertainties that are observed in data for the
four analysis channel SRs and their combination. \label{tab:mean_value_OO}}
\small
\begin{tabular}{l
@{\hspace{4em}}
S[table-number-alignment=right, table-figures-integer=1, table-figures-decimal=2]@{$\,\pm\,$}
S[table-number-alignment=left, table-figures-integer=1, table-figures-decimal=2]
}
\toprule
Channel  & \multicolumn{2}{c}{ \hspace{-3em}$\langle\text{\OOtext}\rangle$} \\
\midrule
\tll~SF  &  -0.54 & 0.72       \\
\tll~DF  &   0.71 & 0.81       \\
\tlhad   &   0.74 & 0.78       \\
\thadhad &  -1.13 & 0.65       \\
\midrule
Combined &  -0.19 & 0.37       \\
\bottomrule
\end{tabular}
\end{table}

To extract confidence intervals for the CP-mixing parameter \dtilde, the ML-fit
described in Section~\ref{sec:fitmodel} is carried out.
The post-fit distributions of the \OOtext~in the various analysis channels are shown in
Figure~\ref{fig:SR_OptObs}. Here the value of the parameter of interest \dtilde, the values of
the nuisance parameters, and the normalizations of the signal and background processes
have been adjusted within their allowed constraints to minimize the NLL curve.  Values of
the NLL are evaluated in steps of $\Delta\dtilde = 0.01$, and the smallest value of the NLL
is observed at $\dtilde = -0.01$. This is the value of \dtilde~that is used for
the post-fit distributions and event yields.  Based upon interpolations between the discrete
evaluations of the various NLL values as a function of \dtilde, the best estimator
for \dtilde~is $-0.013^{+0.048}_{-0.077}$. This value is consistent with the SM expectation of
zero, and no evidence of CPV is observed using this approach.
The best-fit signal strength from the ML-fit is $\mu = 0.73 \pm 0.47$.
 
\begin{figure}[h!]
\centering
\subfloat[]{\includegraphics[width=0.43\linewidth]{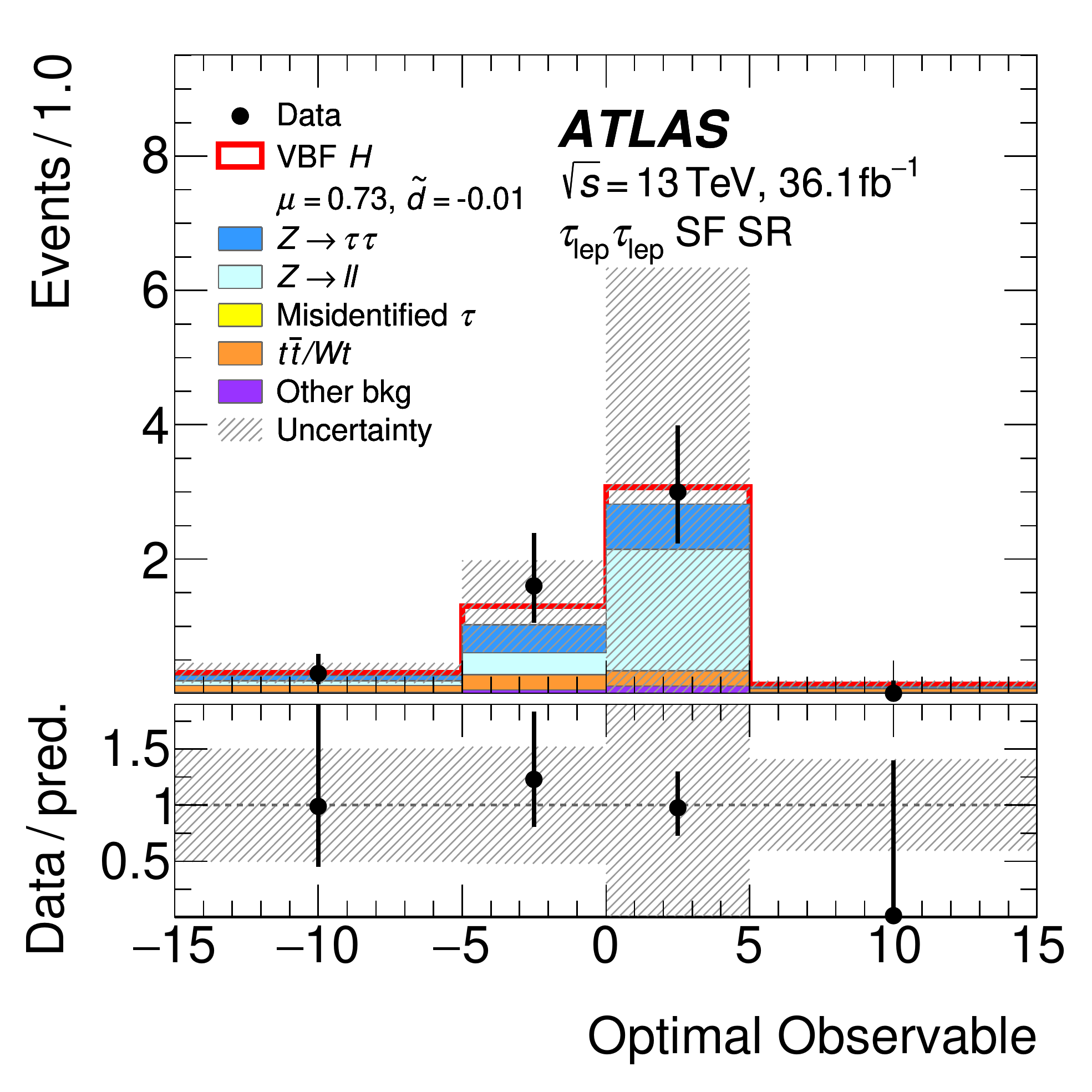} }
\subfloat[]{\includegraphics[width=0.43\linewidth]{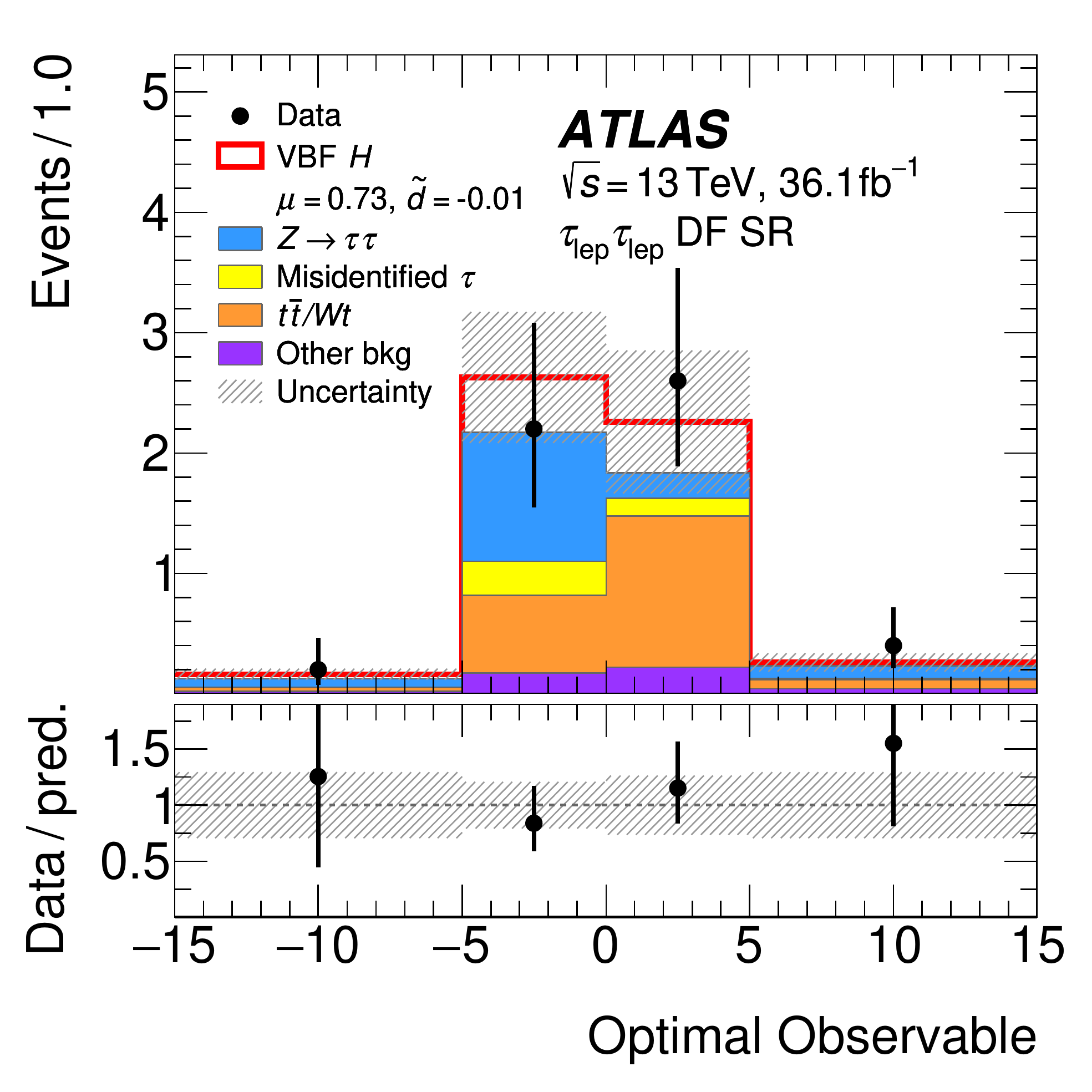} }\\
\subfloat[]{\includegraphics[width=0.43\linewidth]{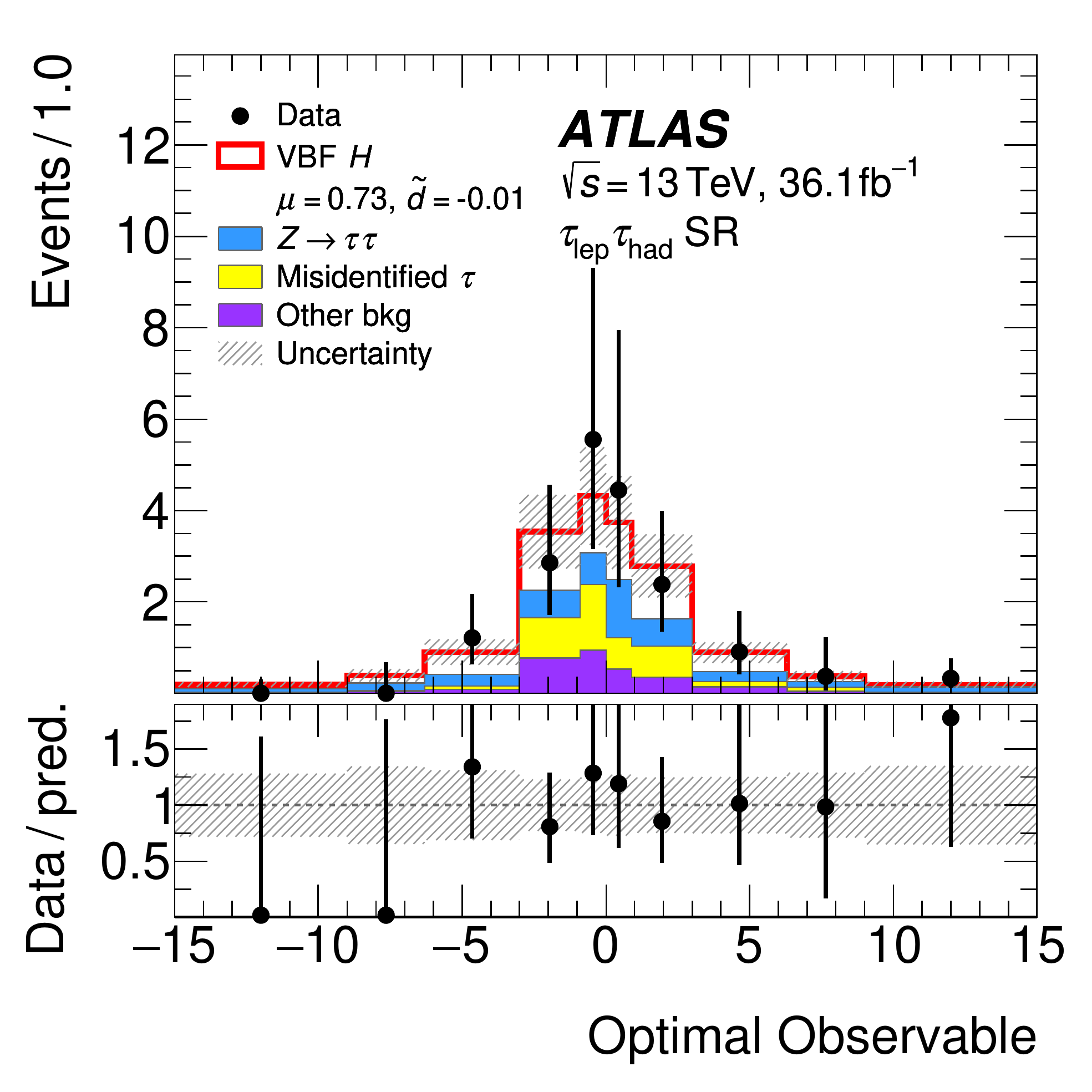} }
\subfloat[]{\includegraphics[width=0.43\linewidth]{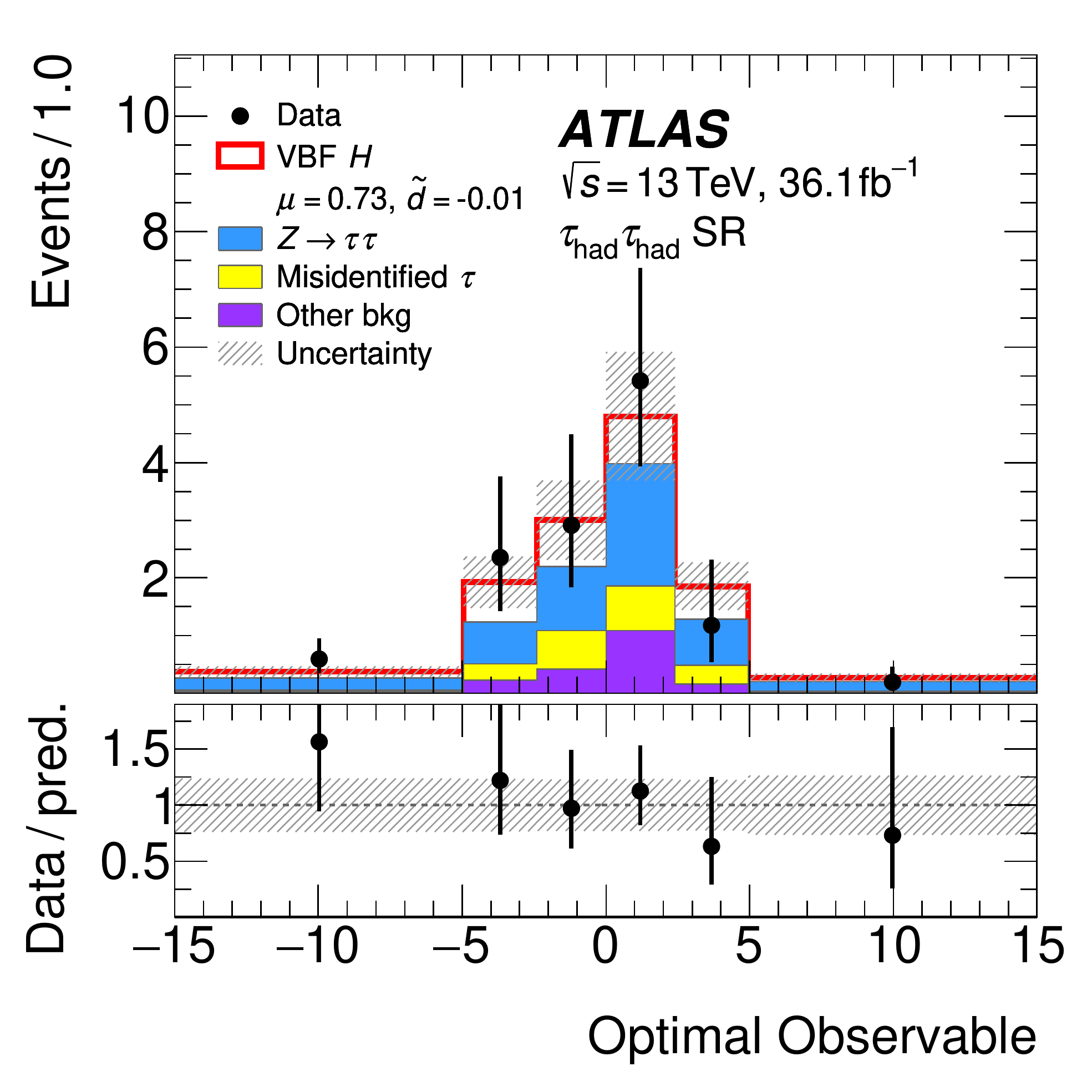} }
\caption{Post-fit distributions of the event yields (divided by the bin width)
as a function of the \OOtext~in the SRs for the
(a)~\tll~SF, (b)~\tll~DF, (c)~\tlhad~and (d)~\thadhad analysis channels.
The values of \dtilde, the signal strength $\mu$, the normalization
of background processes, and nuisance parameters for the event yield prediction
are set to those which minimize the NLL.
The ratios of the data to the prediction are shown in the lower panels.
The size of the combined statistical, experimental and theoretical uncertainties
in the predicted event yields is indicated by the hatched bands.
\label{fig:SR_OptObs}}
\end{figure}

While the predicted background distributions for the \OOtext~are not
perfectly symmetric, they are statistically consistent with a symmetric
distribution. This slight asymmetry causes the expected confidence intervals
for \dtilde~to also be asymmetric.

Tables~\ref{tab:evt_yields_1} and~\ref{tab:evt_yields_2} display
the fitted event yields of the signal ($\mu = 0.73$, $\dtilde = -0.01$) and various
background processes for the SRs
of each channel, along with the corresponding number of events observed in data.
For reference, the signal yields for the SM expectation ($\mu = 1$, $\dtilde = 0$)
are also shown.

\begin{table}[b!]
\centering
\caption{Post-fit event yields in the SRs for the \tll~SF and  \tll~DF analysis channels.
The \Zll~and diboson backgrounds are grouped together in a single background category
for the \tll~DF channel. For comparison, the expected signal
yields for the SM expectation ($\mu = 1$, $\dtilde = 0$) are also shown.\label{tab:evt_yields_1}}
\small
\begin{tabular}{l
S[table-number-alignment=right, table-figures-integer=2, table-figures-decimal=2]@{$\,\pm\,$}
S[table-number-alignment=left, table-figures-integer=2, table-figures-decimal=2]
S[table-number-alignment=right, table-figures-integer=2, table-figures-decimal=2]@{$\,\pm\,$}
S[table-number-alignment=left, table-figures-integer=1, table-figures-decimal=1]
}
\toprule
Process                                       & \multicolumn{2}{c}{\tll~SF}        & \multicolumn{2}{c}{\tll~DF}                 \\
\midrule
Data                                            & \multicolumn{2}{l}{26}             & \multicolumn{2}{l}{30}                       \\
\midrule
VBF $\Htt/WW$ ($\mu = 0.73$, $\dtilde = -0.01$) & 3.3                         & 2.1  & 5.1                  & 3.1                  \\
VBF $\Htt/WW$ ($\mu = 1$, $\dtilde = 0$)        & 4.5                         & 2.9  & 6.9                  & 4.4                  \\
\midrule
\Ztt                                            & 6.6                         & 3.7  & 8.2                  & 3.8                  \\
Fake lepton                                     & 0.02                        & 0.20 & 2.3                  & 0.7                  \\
$t\bar{t}$ + single top                         & 3.8                         & 2.3  & 10.6                 & 5.5                  \\
$\Zll$                                          & 11                          & 18   & \multicolumn{2}{c}{ \multirow{2}{*}{ \hspace{2.0pt}$1.8\enspace\pm\!1.1$ } } \\
Diboson                                         & 0.70                        & 0.59 & \multicolumn{2}{c}{}                      \\
ggF $H$ / $VH$ / $t\bar{t}H$, $\Htt/WW$                   & 0.49                        & 0.48 & 0.70                  & 0.30                  \\
\midrule
Sum of backgrounds                              & 23                          & 17   & 23.6                 & 6.1                    \\
\bottomrule
\end{tabular}
 
\end{table}

\begin{table}[h!]
\centering
\caption{Post-fit event yields in the SRs for the \tlhad~and \thadhad~analysis channels. The line
``Other backgrounds'' includes top quark ($\ttbar$ and single top), diboson, and \Zll~backgrounds.
Backgrounds from $W(\rightarrow \tau_\mathrm{had}\nu)$+jets production in the \thadhad~channel
are also included in ``Other backgrounds''. For comparison, the expected signal
yields for the SM expectation ($\mu = 1$, $\dtilde = 0$) are also shown.
\label{tab:evt_yields_2}}
\small
\begin{tabular}{l
S[table-number-alignment=right, table-figures-integer=2, table-figures-decimal=1]@{$\,\pm\,$}
S[table-number-alignment=left, table-figures-integer=2, table-figures-decimal=1]
S[table-number-alignment=right, table-figures-integer=2, table-figures-decimal=1]@{$\,\pm\,$}
S[table-number-alignment=left, table-figures-integer=1, table-figures-decimal=1]
}
\toprule
Process & \multicolumn{2}{c}{\tlhad} & \multicolumn{2}{c}{\thadhad} \\
\midrule
 
Data & \multicolumn{2}{l}{30} & \multicolumn{2}{l}{37} \\
\midrule
VBF $\Htt$ ($\mu = 0.73$, $\dtilde = -0.01$) & 11.8 & 7.4 & 8.9 & 5.6 \\
VBF $\Htt$ ($\mu = 1$, $\dtilde = 0$)  & 16 & 10 & 12.3 & 7.7 \\
\midrule
\Ztt   & 7.8 & 3.5 & 15.5 & 5.2 \\
Fake lepton/$\tau$ & 6.2 & 1.0 & 5.4 & 2.7 \\
ggF $H$ / $VH$ / $t\bar{t}H$, $\Htt$   & 2.1 & 1.5 & 2.8 & 1.4 \\
Other backgrounds & 2.8 & 3.1 & 2.3 & 0.8 \\
\midrule
Sum of backgrounds  & 19.0 & 5.5 & 26.0 & 6.6 \\
\bottomrule
\end{tabular}
\end{table}

The observed and expected $\Delta\mathrm{NLL}$ curves are shown in
Figure~\ref{fig:NLL_curves}(a) as a function of \dtilde.
The expected curves are obtained in a two-step process: firstly,
nuisance parameters and background normalization factors are constrained via a ML-fit to all
analysis CRs, excluding the SRs;
then another fit is performed in all SRs and CRs to pseudo-data which were
created with the best-fit parameter values from the first step.
This two-step process ensures that the nuisance parameters
and the background normalization factors for the expected sensitivity are set to values
that are consistent with the observed data in the analysis CRs. The expected $\Delta\mathrm{NLL}$ curve
is shown for $\dtilde=0$ and $\mu=1$ and represents the best estimate
of the sensitivity of the analysis based on SM expectations. Another $\Delta\mathrm{NLL}$ curve
with $\dtilde=0$ and the signal strength $\mu$ set to the observed value of $0.73$
is also shown in order to demonstrate the decrease in sensitivity due to the lower than
expected event yields (see Tables~\ref{tab:evt_yields_1} and~\ref{tab:evt_yields_2}).
Also shown for comparison in Figure~\ref{fig:NLL_curves}(a) is the pre-fit expected
$\Delta\mathrm{NLL}$ curve, which is obtained
using a pseudo-dataset where the event yields and distributions in the SRs and CRs are set
to the SM expectations for both the signal (with $\dtilde=0$ and $\mu=1$) and background processes.
This demonstrates that the
preferred values of the nuisance parameters and normalization factors based on the observed
data in the background CRs in the expected $\Delta\mathrm{NLL}$ curve
result in a decrease in sensitivity to \dtilde~when compared with the pre-fit expected
curve.

\begin{figure}[h!b!t!]
\centering
\subfloat[]{\includegraphics[width=0.43\linewidth]{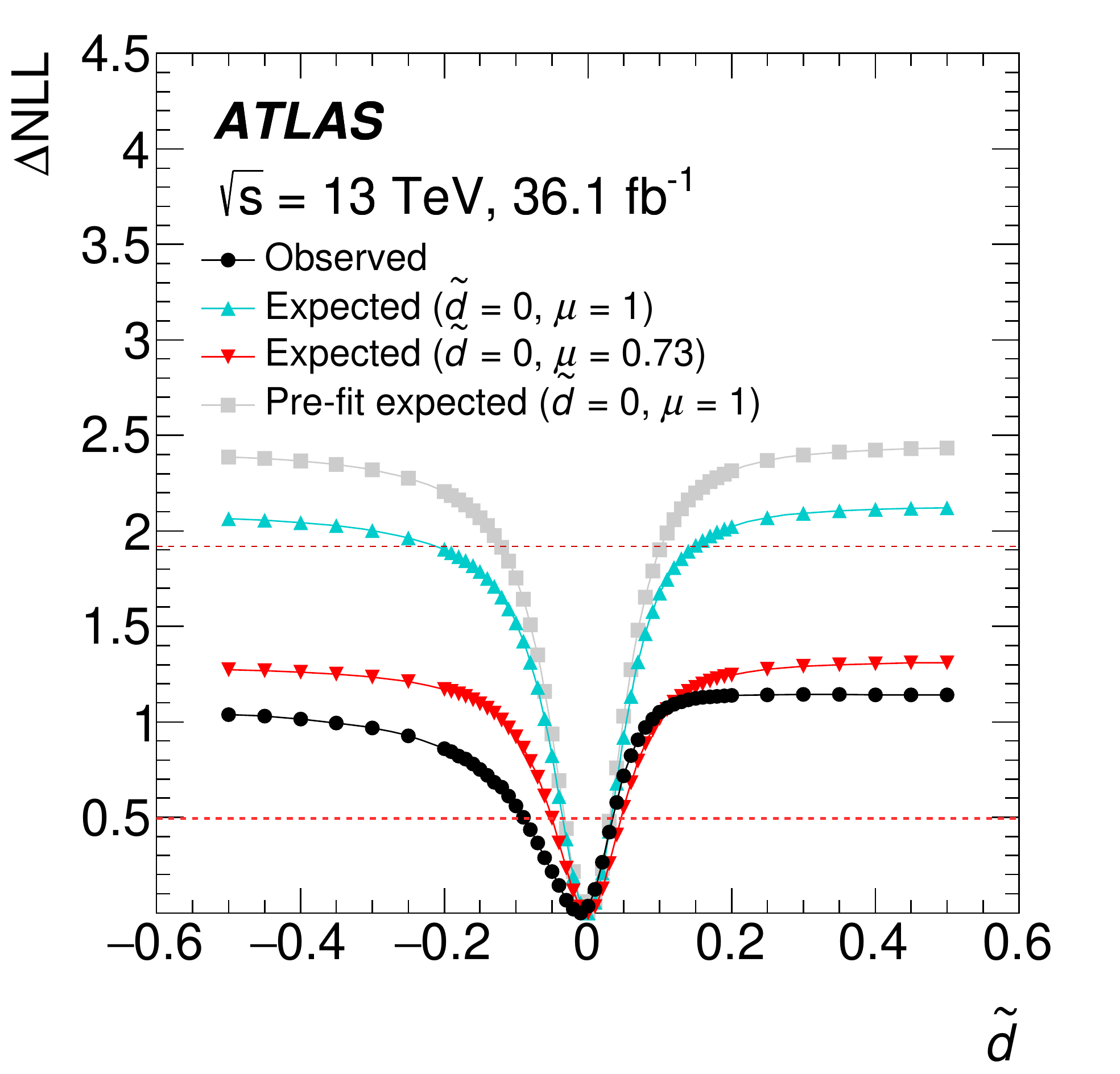} }
\subfloat[]{\includegraphics[width=0.43\linewidth]{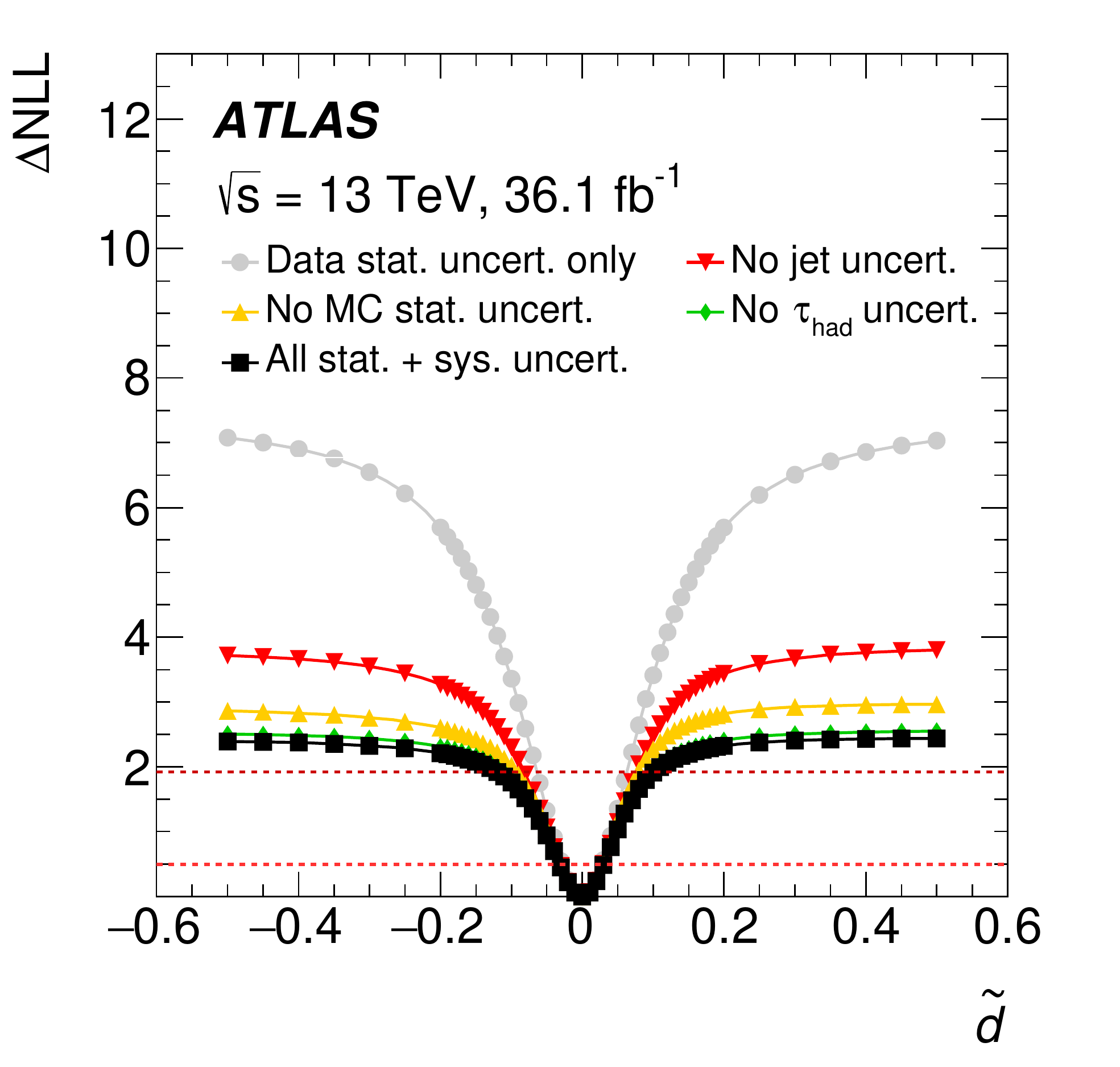} } \\
\subfloat[]{\includegraphics[width=0.43\linewidth]{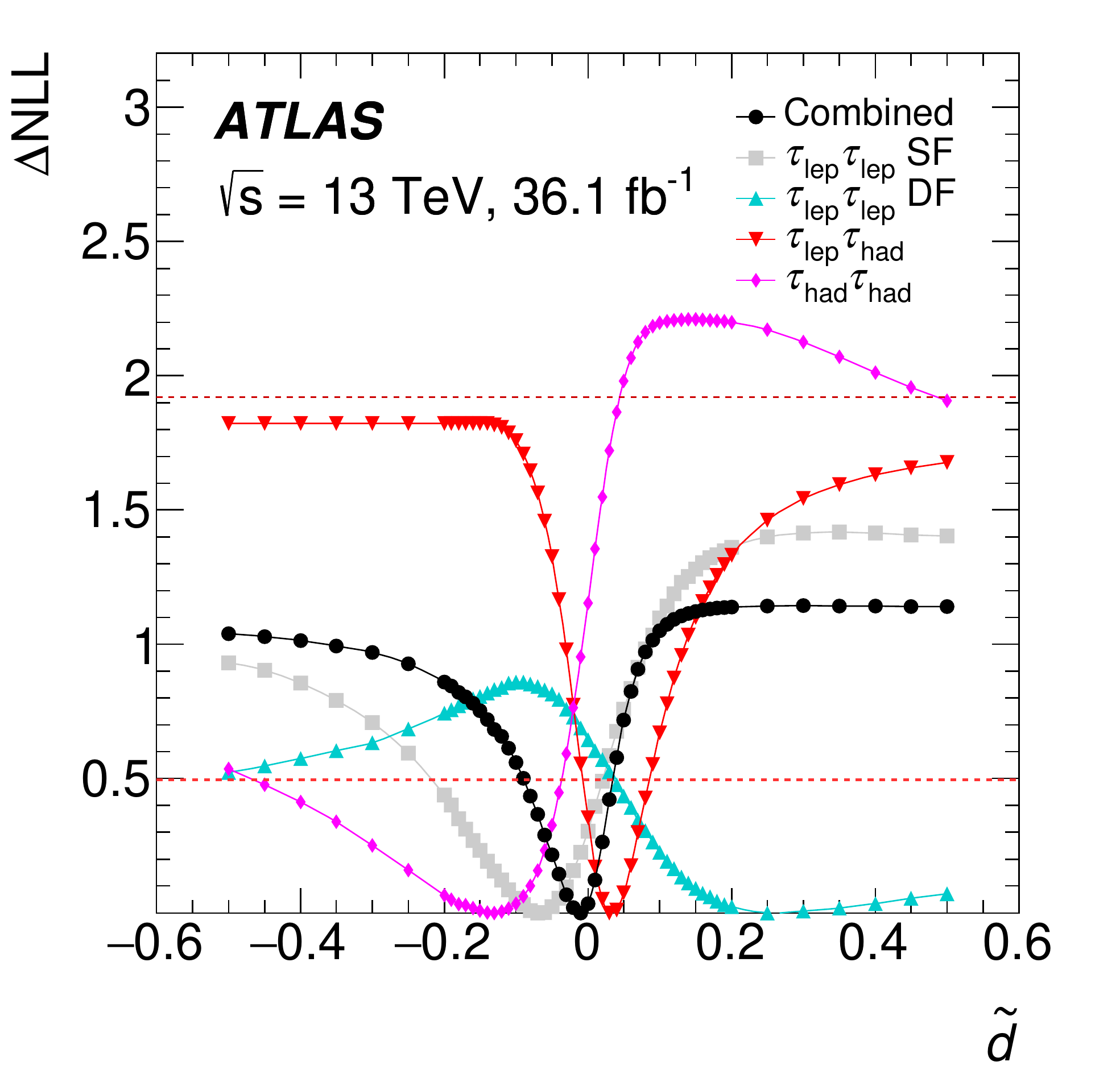} }
\caption{(a) The observed $\Delta\mathrm{NLL}$ curve as a function of \dtilde~values.
For comparison, expected $\Delta\mathrm{NLL}$ curves are also shown.
The constraints on the nuisance parameters and normalization factors are
first determined in a CR-only fit, and then a fit to pseudo-data corresponding
to these nuisance parameters, normalization factors, and to
$\dtilde=0, \mu = 1$ or $\dtilde=0, \mu = 0.73$ is performed to obtain
these $\Delta\mathrm{NLL}$ curves. A pre-fit expected $\Delta\mathrm{NLL}$
is also shown, using pseudo-data corresponding to $\dtilde=0$ and $\mu = 1$ in the signal
and control regions.
(b) The expected $\Delta\mathrm{NLL}$ curves ($\dtilde=0, \mu = 1$) comparing
the sensitivity of the fit with and without systematic uncertainties.
For comparison, other curves are shown which remove the effect of jet-based
systematic uncertainties, $\tau$-based systematic uncertainties,
and MC statistical uncertainties.
(c) The observed $\Delta\mathrm{NLL}$ curves for each analysis channel as a function of
\dtilde, compared with the combined result. For the individual analysis channel
$\Delta\mathrm{NLL}$ curves, only event yield information in the other SRs is used,
ensuring that the \OOtext~distributions in the other SRs do not influence the
preferred value of \dtilde. The signal strength is constrained to be positive
in these individual channel $\Delta\mathrm{NLL}$ curves.
For all figures, the dashed horizontal lines show the values of $\Delta\mathrm{NLL}$
used to define the 68\% and 95\% confidence intervals.
\label{fig:NLL_curves}}
\end{figure}
 
The effect of systematic uncertainties on the sensitivity to \dtilde~can be seen in
Figure~\ref{fig:NLL_curves}(b). Here, the expected $\Delta\mathrm{NLL}$ curves are shown
for $\dtilde=0$ and $\mu = 1$, with and without the effect of systematic uncertainties. To assess the
impact of systematic uncertainties stemming from jet reconstruction,
$\tau$-lepton identification,
and MC sample size, expected $\Delta\mathrm{NLL}$ curves are also shown
where the nuisance parameters corresponding to the systematic uncertainties in question
have been removed from the likelihood function. It is evident that the experimental
uncertainties related to jet reconstruction have the largest effect on the sensitivity of
this analysis to \dtilde.
 
To obtain insight into the preferred values of \dtilde~obtained for the individual
\OOtext~distributions in the different analysis channels,  $\Delta\mathrm{NLL}$
curves for each individual channel are shown in Figure~\ref{fig:NLL_curves}(c),
and compared with the combined result. For these individual $\Delta\mathrm{NLL}$ curves,
only event yield information from the other three signal regions that are not being
featured is used, so that the  distribution of events in the \OOtext~in these other signal
regions is not exploited in the ML-fit. For these individual channel $\Delta\mathrm{NLL}$
curves, the signal strength is constrained to be positive so that the ML-fit is stable
and insensitive to event yield fluctuations in the individual channel SRs that arise
from smaller samples. This constraint is responsible for the plateau in the $\Delta\mathrm{NLL}$
curve occurring at negative values in the \tlhad~channel.

An observed 68\% CL interval of $\dtilde \in [-0.090,0.035]$
is obtained from the observed $\Delta\mathrm{NLL}$ curve using \OOtext~distributions in all
SRs. The corresponding expected interval, based upon the expected $\Delta\mathrm{NLL}$ curve
for  $\dtilde = 0$ and $\mu=1$  in
Figure~\ref{fig:NLL_curves}(a) is $\dtilde \in [-0.035,0.033]$.
This represents an improvement on the confidence interval
for \dtilde~set in Ref.~\cite{HIGG-2015-06}.
While no observed 95\% CL interval for \dtilde~can be quoted, the
corresponding expected interval is $\dtilde \in [-0.21,0.15]$ at 95\% CL.
The asymmetry in these expected intervals stems from the slightly asymmetric
\OOtext~distribution of the background estimates in the SRs, caused by the limited sample
sizes for the background predictions.

The intervals based upon the
pre-fit expected $\Delta\mathrm{NLL}$ curve in Figure~\ref{fig:NLL_curves}(a),
where the nuisance parameters and background normalization factors do not take into
account the observed data in the CRs, are $\dtilde \in [-0.032,0.031]$ at 68\% CL
and $\dtilde \in [-0.12,0.10]$  at 95\% CL.

 
\FloatBarrier

\section{Conclusion}
 
The CP invariance of the Higgs boson coupling to vector bosons has been tested
in the VBF $\Htt$ process in 36.1~fb$^{-1}$ of $\sqrt{s}=13$~\TeV\ proton--proton
collision data obtained with the ATLAS detector at the LHC. In this analysis, an
\OOtext~was used and confidence intervals were
set on the CP-mixing parameter \dtilde.
 
Since the mean of the \OOtext~observed in data is consistent with zero,
and the obtained confidence intervals for \dtilde~are consistent
with the Standard Model value $\dtilde=0$,
no evidence of CP violation is observed from this analysis.
Due to lower than expected signal yields in data, no constraints on \dtilde~can be set at
95\%~CL, while the corresponding Standard Model expectation
is $\dtilde \in [-0.21,0.15]$.
An observed 68\%~CL interval of
$\dtilde \in [-0.090,0.035]$ is obtained, while the corresponding interval
based on the expectation is $\dtilde \in [-0.035,0.033]$.
 
 
\section*{Acknowledgements}
 

We thank CERN for the very successful operation of the LHC, as well as the
support staff from our institutions without whom ATLAS could not be
operated efficiently.
 
We acknowledge the support of ANPCyT, Argentina; YerPhI, Armenia; ARC, Australia; BMWFW and FWF, Austria; ANAS, Azerbaijan; SSTC, Belarus; CNPq and FAPESP, Brazil; NSERC, NRC and CFI, Canada; CERN; CONICYT, Chile; CAS, MOST and NSFC, China; COLCIENCIAS, Colombia; MSMT CR, MPO CR and VSC CR, Czech Republic; DNRF and DNSRC, Denmark; IN2P3-CNRS and CEA-DRF/IRFU, France; SRNSFG, Georgia; BMBF, HGF and MPG, Germany; GSRT, Greece; RGC and Hong Kong SAR, China; ISF and Benoziyo Center, Israel; INFN, Italy; MEXT and JSPS, Japan; CNRST, Morocco; NWO, Netherlands; RCN, Norway; MNiSW and NCN, Poland; FCT, Portugal; MNE/IFA, Romania; MES of Russia and NRC KI, Russia Federation; JINR; MESTD, Serbia; MSSR, Slovakia; ARRS and MIZ\v{S}, Slovenia; DST/NRF, South Africa; MINECO, Spain; SRC and Wallenberg Foundation, Sweden; SERI, SNSF and Cantons of Bern and Geneva, Switzerland; MOST, Taiwan; TAEK, Turkey; STFC, United Kingdom; DOE and NSF, United States of America. In addition, individual groups and members have received support from BCKDF, CANARIE, Compute Canada and CRC, Canada; ERC, ERDF, Horizon 2020, Marie Sk{\l}odowska-Curie Actions and COST, European Union; Investissements d'Avenir Labex, Investissements d'Avenir Idex and ANR, France; DFG and AvH Foundation, Germany; Herakleitos, Thales and Aristeia programmes co-financed by EU-ESF and the Greek NSRF, Greece; BSF-NSF and GIF, Israel; CERCA Programme Generalitat de Catalunya and PROMETEO Programme Generalitat Valenciana, Spain; G\"{o}ran Gustafssons Stiftelse, Sweden; The Royal Society and Leverhulme Trust, United Kingdom.
 
The crucial computing support from all WLCG partners is acknowledged gratefully, in particular from CERN, the ATLAS Tier-1 facilities at TRIUMF (Canada), NDGF (Denmark, Norway, Sweden), CC-IN2P3 (France), KIT/GridKA (Germany), INFN-CNAF (Italy), NL-T1 (Netherlands), PIC (Spain), ASGC (Taiwan), RAL (UK) and BNL (USA), the Tier-2 facilities worldwide and large non-WLCG resource providers. Major contributors of computing resources are listed in Ref.~\cite{ATL-GEN-PUB-2016-002}.
 
 
\clearpage
\appendix
\printbibliography
 
\clearpage 
 
\begin{flushleft}
\hypersetup{urlcolor=black}
{\Large The ATLAS Collaboration}

\bigskip

\AtlasOrcid[0000-0002-6665-4934]{G.~Aad}$^\textrm{\scriptsize 102}$,    
\AtlasOrcid[0000-0002-5888-2734]{B.~Abbott}$^\textrm{\scriptsize 129}$,    
\AtlasOrcid{D.C.~Abbott}$^\textrm{\scriptsize 103}$,    
\AtlasOrcid[0000-0002-2788-3822]{A.~Abed~Abud}$^\textrm{\scriptsize 36}$,    
\AtlasOrcid[0000-0002-1002-1652]{K.~Abeling}$^\textrm{\scriptsize 53}$,    
\AtlasOrcid[0000-0002-2987-4006]{D.K.~Abhayasinghe}$^\textrm{\scriptsize 94}$,    
\AtlasOrcid[0000-0002-8496-9294]{S.H.~Abidi}$^\textrm{\scriptsize 167}$,    
\AtlasOrcid[0000-0002-8279-9324]{O.S.~AbouZeid}$^\textrm{\scriptsize 40}$,    
\AtlasOrcid{N.L.~Abraham}$^\textrm{\scriptsize 156}$,    
\AtlasOrcid[0000-0001-5329-6640]{H.~Abramowicz}$^\textrm{\scriptsize 161}$
% ...
\bigskip
\\

% ...
$^{36}$CERN, Geneva; Switzerland.\\
% ...
$^{40}$Niels Bohr Institute, University of Copenhagen, Copenhagen; Denmark.\\
% ...
$^{53}$II. Physikalisches Institut, Georg-August-Universit\"{a}t G\"ottingen, G\"ottingen; Germany.\\
% ...
$^{94}$Department of Physics, Royal Holloway University of London, Egham; United Kingdom.\\
% ...
$^{102}$CPPM, Aix-Marseille Universit\'e, CNRS/IN2P3, Marseille; France.\\
$^{103}$Department of Physics, University of Massachusetts, Amherst MA; United States of America.\\
% ...
$^{129}$Homer L. Dodge Department of Physics and Astronomy, University of Oklahoma, Norman OK; United States of America.\\
% ...
$^{156}$Department of Physics and Astronomy, University of Sussex, Brighton; United Kingdom.\\
% ...
$^{161}$Raymond and Beverly Sackler School of Physics and Astronomy, Tel Aviv University, Tel Aviv; Israel.\\
% ...
$^{167}$Department of Physics, University of Toronto, Toronto ON; Canada.\\
% ...

\end{flushleft}


\end{document}